\documentclass[aps,amsmath,amssymb,pr,twocolumn]{revtex4-1}
\usepackage{bm}
\usepackage{epsfig}
\usepackage{amsmath}
\usepackage{amsmath}
\usepackage{ulem}
\newcommand{\nix}[1]{}
\usepackage[unicode=true,colorlinks=true,urlcolor=blue,citecolor=blue]{hyperref}

\begin{document}

\title{Circular and magnetoinduced photocurrents in Weyl semimetals}

\author{L.\,E.\,Golub and E.\,L.\,Ivchenko}
\affiliation{Ioffe Institute, St.~Petersburg 194021, Russia}


\begin{abstract}
We develop a theory of the direct interband and indirect intraband photogalvanic effects in Weyl semimetals belonging to the gyrotropic classes with improper symmetry operations. At zero magnetic field, an excitation of such a material with circularly polarized light leads to a photocurrent whose direction depends on the light helicity. We show that in the semimetals of the C$_{2v}$ symmetry, an allowance for the tilt term in the effective Hamiltonian is enough to prevent cancellation of the photocurrent contributions from the Weyl cones of opposite chiralities. In the case of the 
C$_{4v}$ symmetry, in addition to the tilt it is necessary  to include terms of the second- or third-order in the electron quasi-momentum. For indirect intraband transitions, the helicity-dependent photocurrent generated within each Weyl node takes on a universal value determined by the fundamental constants, the light frequency and electric field. We have complementarily investigated the magneto-gyrotropic photogalvanic effect, i.e.  an appearance of a photocurrent under unpolarized excitation in a magnetic field. In quantized magnetic fields, the photocurrent is caused by optical transitions between the one-dimensional magnetic subbands. A value of the photocurrent is particularly high if  one of the photocarriers is excited to the chiral subband with the energy below the cyclotron energy.

\end{abstract}

\maketitle

\section{Introduction}
Weyl semimetals (WSMs) represent materials allowing for a solid-state realization of three-dimensional massless fermions. After discovery of a nonvanishing neutrino mass, the Weyl fermions in condensed matter systems are unique particles with a massless linear energy dispersion and a definite chirality. This determines remarkable properties of WSMs, for a recent review see Ref.~\cite{RMP_2018}. A unique feature is the chiral anomaly consisting in a nonconservation of the particle number of a given chirality and leading to a negative magnetoresistance. Another important fact is that each Weyl node serves as a magnetic monopole in the reciprocal space with a nonzero Berry curvature. The important consequence is an existence of topologically induced Fermi arcs at a surface of a Weyl semimetal. Due to  topological reasons, each Weyl node has its counterpart of opposite chirality. In contrast to neutrinos, the dispersion cones can be tilted or even overtilted which is realized, respectively, in type~I and type~II WSMs \cite{RMP_2018}.
 
The discovery of WSMs has been followed by theoretical and experimental studies of their transport and optical properties, both linear and nonlinear. In particular, it has been shown that the interaction of circularly-polarized light with chiral fermions is governed by the Berry curvature of the Weyl node. This allows a new look at the Circular PhotoGalvanic Effect (CPGE), an appearance of a helicity-dependent electric photocurrent upon illumination of the sample by circularly polarized light~\cite{EL_book}. The CPGE is allowed by the symmetry of gyrotropic (or optically active) media, and many WSMs belong to the family of gyrotropic crystals. It has been demonstrated that, in each Weyl node, the CPGE  has universal features independent of details of the material. In particular, the generation rate of the CPGE current density is determined, except for the 
intensity of the exciting light, by fundamental constants~\cite{Moore}. However, the corresponding CPGE current has the opposite polarity in the Weyl node of opposite chirality. Since the Weyl nodes exist in pairs of opposite chirality, the universality of CPGE is preserved only in those semimetals where such Weyl nodes have different energies. This is possible in the absence of any improper symmetry operations because such an operation transforms a given Weyl node to another one of the opposite chirality. In contrast, the point-group symmetry of available WSMs contains improper operations, and the universal contributions of individual nodes to the CPGE current mutually compensate each other. The net photocurrent can be generated in real Weyl semimetals owing to corrections to the Weyl Hamiltonian and, hence, to the carrier energy dispersion. It has been shown~\cite{Patrick} that the net CPGE current induced within two Weyl nodes of opposite chirality becomes nonvanishing in tilted WSMs where it, however, loses its universality and depends on the tilt. As we pointed out~\cite{JETP_Lett_2017}, the net CPGE current is sensitive to the sign of parameters describing corrections to the Weyl Hamiltonian rather than to chirality of the Weyl fermions. This opens a possibility to study the real Hamiltonians in WSMs with the help of CPGE. The theoretical study is also stimulated by the recent observations of the CPGE in the TaAs semimetal~\cite{Gedik,China}.

In the present paper we consider WSMs of the crystal classes containing improper symmetry operations and derive minimal models which allow for the CPGE. For this purpose we include into the electron effective Hamiltonian linear and nonlinear, spin-dependent and spin-independent terms leading to the net CPGE current under direct optical transitions between the valence and conduction band states as well as under intraband indirect transitions. It is demonstrated that the intraband CPGE current induced in an individual Weyl node has a universal value independent of the momentum relaxation details.

An additional effect specific to gyrotropic media is a Magneto-gyrotropic PhotoGalvanic Effect (MPGE) which is a generation of photocurrent odd in the magnetic field and independent of the light polarization. It has been investigated in bulk semiconductors~\cite{MPGE_bulk1,MPGE_bulk2} and quantum-well structures~\cite{MPGE_QWs}. Recently the MPGE  has been theoretically studied in WSMs in weak magnetic fields and termed the helical magnetic effect~\cite{Kharzeev}. Here we consider gyrotropic WSMs in a quantizing magnetic field and investigate the MPGE current injected by direct optical transitions between magnetic subbands. In multi-valley semimetals of the C$_{2v}$ symmetry with Weyl nodes of opposite chiralities the net MPGE current appears with allowance for the
tilt. 

The paper is organized as follows. In Sect.~\ref{interband}, we study the CPGE in semimetals of the C$_{2v}$ and C$_{4v}$ point-group symmetries at zero magnetic field. We take into account both the tilt and spin-dependent nonlinear terms in the effective Hamiltonian. The photocurrent generated under intraband indirect optical absorption is calculated in Section~\ref{intraband}. In Sect.~\ref{Sec_MPGE}, we consider the MPGE realized  under the unpolarized photoexcitation in an external magnetic field. In Sect.~\ref{disc} we present and analyze a typical spectral dependence of the MPGE current and make some comments concerning non-stationary photocurrents and role of the electron-electron interaction. Section~\ref{concl} concludes the paper.

\section{Circular photocurrent at inter-band transitions}

\label{interband}

We start with the effective electron Hamiltonian 
\begin{equation} \label{Hamilt3}
{\cal H} =\sigma_x d_x({\bm k}) + \sigma_y d_y({\bm k}) + \sigma_z d_z({\bm k})  + \sigma_0d_0({\bm k})\:,
 \end{equation}
where $\sigma_{\alpha}$ ($\alpha = x, y, z$) are the Pauli spin matrices, $\sigma_0$ is the unit 2$\times$2 matrix,  ${\bm k}$ is  the electron wavevector referred to a particular Weyl node ${\bm k}_W$ and the functions $d_{\alpha}({\bm k}), d_0({\bm k})$ can be expanded in a Taylor series starting from the first-order terms. The energy dispersion consists of two branches $ E_{\pm, {\bm k}}  = d_0({\bm k}) \pm d({\bm k})$, where $d = |{\bm d}|$.
The photocurrent generated under direct optical transitions in the vicinity of a Weyl node is given by
\begin{equation} \label{general}
{\bm j} = e \sum\limits_{\bm k} {\bm v}_{+-} ({\bm k}) \tau_p W_{+-}({\bm k})\:.
\end{equation}
Here $W_{+-}({\bm k})$ is the transition rate related with the optical matrix element $M_{+-}$ by the Fermi golden rule
 \begin{equation} \label{W+-}
W_{+-} = \frac{2 \pi}{\hbar} \left\vert M_{+-} \right\vert^2 F({\bm k})  \delta \left[ 2d({\bm k}) - \hbar \omega \right]\:,
\end{equation}
$F({\bm k})$ is the difference $f_0(E_{-,{\bm k}}) - f_0(E_{+,{\bm k}})$ of the electron occupations of the initial and final states
with $f_0$ being  the Fermi-Dirac distribution function, $\omega$ is the light wave frequency, $\tau_p$ is the electron and hole momentum relaxation times assumed to coincide, and $ {\bm v}_{+-} ({\bm k})$ is the difference of the group velocities 
\[
{\bm v}_{+-} ({\bm k})= \frac{1}{\hbar} \frac{\partial }{\partial {\bm k}} \left( E_{+, {\bm k}}  - E_{-, {\bm k}} \right)=
\frac{2}{\hbar} \frac{\partial d({\bm k})}{\partial {\bm k}}\:.
\]
A special property of WSMs is the following relation between the contribution to $\left\vert M_{+-} \right\vert^2$ dependent on the light circular polarization and the Berry curvature ${\bm \Omega_{\bm k}}$ \cite{JETP_Lett_2017}
\begin{equation} \label{MOmega}
\left\vert  M_{+-} \right\vert^2_{\rm circ} = |{\bm E}|^2  \frac{2 e^2 d^2}{(\hbar \omega)^2} {\bm \varkappa}\cdot{\bm \Omega_{\bm k}} \:,
\end{equation}
where ${\bm E}$ is the amplitude of the light's electric field and ${\bm \varkappa} = i({\bm E} \times {\bm E}^*)/|{\bm E}|^2$ is the photon helicity. For the transverse electromagnetic wave, ${\bm \varkappa}$ is expressed via the light wave vector ${\bm q}$ and the degree of circular polarization $P_\text{circ}$ as  $\bm \varkappa =P_\text{circ} {\bm q}/q$. The Berry curvature is given in terms of the vector ${\bm d}(\bm k)$ as follows
\begin{equation} 
\label{Berrygeneral}
 \Omega_{{\bm k},i} = \frac{\bm d }{2d^3} \cdot  
 \left( \frac{\partial \bm d }{\partial k_{i+1}} \times \frac{\partial \bm d}{\partial k_{i+2}} \right) \:,
\end{equation}
where the cyclic permutation of indices is assumed.

\subsection{Linear spin-orbit coupling}
\label{Linear_SOC}

If the spin-dependent part of the Hamiltonian (\ref{Hamilt3}) is linear in ${\bm k}$ and the tilt term vanishes then the circular photocurrent takes the universal form 
\begin{equation} \label{Gamma00}
{\bm j} = {\cal C} \Gamma_0\tau_p i \left( {\bm E} \times {\bm E}^* \right)\:.
\end{equation}
Here $\Gamma_0 = \pi e^3 /3 h^2$,  $e$ is the electron charge, $h$  is the
Planck constant, and ${\cal C} = \pm 1$ is the chirality (or topological charge) of the Weyl node
\begin{equation}
{\cal C} = {1\over 2\pi}\int\limits_\Sigma \bm \Omega_{\bm k} \cdot d\bm S_{\bm k},
\end{equation}
where the integration performed over a closed surface $\Sigma$ with the Weyl point inside.

The presence of crystal symmetry operations transforming the Weyl node ${\bm k}_W$ to another point of the Brillouin zone determines a multi-valley character of the WSM electron band structure. If the gyrotropic crystal class does not contain refection planes then the circular photocurrent related to the node ${\bm k}_W$ is given by Eq.~(\ref{Gamma00}) times the number of valleys and retains the universality. However, if the gyrotropic class contains a reflection plane $\sigma$ (or a rotoinversion axis $S_n$) then the nodes ${\bm k}_W$ and $\sigma{\bm k}_W$ (or $S_n\bm k_W$) are characterized by opposite chiralities, and their contributions to ${\bm j}$ can compensate each other, partly or completely. Let us analyze how the presence of
reflection planes in the point-symmetry group ${\cal F}$ of a gyrotropic crystal affects the CPGE
if the Hamiltonian has the form
\begin{equation} \label{lineartilt}
{\cal H} = {\cal C}\hbar v_0 \bm \sigma \cdot \bm k + \sigma_0 d_0(\bm k)\:,
\end{equation}
where $v_0>0$ is the electron effective velocity, and the tilt $d_0(\bm k)$ is an analytical function of ${\bm k}$ vanishing at the point ${\bm k}_W$. For this purpose it is useful to consider the one-dimensional representation ${\cal D}_v$ of the group ${\cal F}$ defined as follows: ${\cal D}_v(g) = {\cal C}_g/{\cal C}$, where ${\cal C}$ and ${\cal C}_g$ are the chiralities of the fixed node ${\bm k}_W$ and  the node $g {\bm k}_W$. For the Hamiltonian (\ref{lineartilt}) one has ${\bm v}_{+-} ({\bm k}) = 2v_0{\bm k}/k$ and ${\bm \varkappa}\cdot{\bm \Omega_{\bm k}} \propto {\cal C} {\bm \varkappa}\cdot{\bm k}/2k^3$. Therefore the photocurrent $j_{\alpha}^{\beta} \propto {\varkappa}_{\beta}$ summed up over all the valleys $g {\bm k}_W$ can be presented as
\[
j_{\alpha}^{\beta} = {\cal C}\sum\limits_{\bm k} Q(k) k_{\alpha} k_{\beta} \sum_{g \in {\cal F}} {\cal D}_v(g) F\left( g^{-1} {\bm k}\right)\:,
\]
where $Q(k)$ is a function of the modulus $k$. By the variable transformation ${\bm k} \to g {\bm k}$ the sum is reduced to
\begin{equation} \label{kgk}
j_{\alpha}^{\beta} = {\cal C} \sum\limits_{\bm k} Q(k)  F({\bm k}) \sum_{g \in {\cal F}} {\cal D}_v(g) \: g \left( k_{\alpha} k_{\beta} \right) \:.
\end{equation}
Thus, denoting by ${\cal D}^{(1)}$ the three-dimensional representation according to which the wavevector components $k_{\alpha}$ transform, we can formulate a theorem:  among $j_{\alpha}^{\beta}$ there are nonzero values only if the symmetrized direct product $\left[{\cal D}^{(1)} \times {\cal D}^{(1)}\right]$ contains the representation ${\cal D}_v$. 

For the point group C$_{2v}$ with the reflection planes $\sigma_v(xz)$ and $\sigma_v(yz)$, the representation ${\cal D}_v$ coincides with the representation  $A_2$. Among linear combinations of the functions $k_{\alpha} k_{\beta}$ only the product $k_x k_y$ transforms according to $A_2$. It follows then that the sum in Eq.~(\ref{kgk}) is nonzero only for the pairs $\alpha = x, \beta = y$ and $\alpha = y, \beta = x$.  With allowance for the fraction of $F({\bm k})$ of the $A_2$ symmetry the photocurrent off-diagonal components $j_x^y = j_y^x$ become nonzero. However, the antisymmetric contribution $(j_x^y - j_y^x)/2$
to the photocurrent also allowed by the C$_{2v}$ point group is absent in the model (\ref{lineartilt}). Note that in the axes $x',y'$ rotated around $z$ by 45$^{\circ}$ with respect to the $x,y$ axes the model (\ref{lineartilt}) allows the diagonal components $j_{x'}^{x'} = - j_{y'}^{y'}$.

For the point group C$_{4v}$, the representation ${\cal D}_v$ coincides with the representation denoted also by $A_2$ (or $\Gamma_2$ in the other notation). In this group, however, the symmetrized product $\left[{\cal D}^{(1)} \times {\cal D}^{(1)}\right]$ is decomposed into irreducible representations $2 A_1 + B_1 + B_2 + E$ (or $2 \Gamma_1 + \Gamma_3 + \Gamma_4 + \Gamma_5$) and does not contain the representation $A_2$. All the components $j_{\alpha}^{\beta}$ vanish for the linear spin-orbit coupling and an arbitrary tilt function $d_0({\bm k})$. Thus, we eventually come to conclusion that the CPGE in the C$_{4v}$ symmetry crystals, as well as the difference between $j_x^y$ and $j_y^x$ (or the components $j_{x'}^{y'} = - j_{y'}^{x'}$) allowed by the C$_{2v}$ symmetry, can be obtained only  by adding to the spin-dependent part of ${\cal H}$ corrections of the higher order in ${\bm k}$. In the next section we demonstrate how this is done for the C$_{4v}$ point-group symmetry.

\subsection{Account for spin-dependent nonlinear terms}
In order to calculate the CPGE in WSMs of the C$_{4v}$ symmetry we present the functions $d_{\alpha}({\bm k})$  in a more general form 
$${\cal C}\hbar v_0 k_x + P_x({\bm k}), \quad {\cal C}\hbar v_0 k_y + P_y({\bm k}), \quad {\cal C}\hbar v_0 k_z\:,$$ respectively, where the expansion of $P_{\alpha}({\bm k})$ in powers of ${\bm k}$ starts from the second or third order terms. We focus on the calculation of the symmetry allowed component $j_x^y$ given by a sum of the individual contributions 
\begin{equation} \label{C4v}
j_x^y(Wi) = \frac{2\pi e^3 \tau_p }{\hbar^2} |{\bm E}|^2 \sum\limits_{\bm k} \frac{\partial d}{\partial k_x}  \Omega_{{\bm k},y} F({\bm k})\  \delta ( 2d - \hbar \omega )\:,
\end{equation}
where the index $i$ runs from 1 to the number $N$ of the equivalent nodes. The component $j_y^x(Wi)$ differs from Eq.~(\ref{C4v}) only in sign.

We assign number 1 to one of the Weyl nodes ${\bm k}_{W1}$ lying, e.g., in the region of the Brillouin zone with the positive components ${\bm k}_{W1,\alpha} > 0$. Bearing in mind 8 elements of spatial symmetry of the C$_{4v}$ group and the time-inversion symmetry, we have 16 equivalent nodes. To analyze the net electric photocurrent, it is sufficient to consider only two nodes, the fixed one, ${\bm k}_{W1}$, and the node ${\bm k}_{W2}$ obtained by the reflection in the diagonal plane $\sigma_d$ which transforms $k_x, k_y, k_z$ into $k_y, k_x, k_z$ and $\sigma_x, \sigma_y, \sigma_z$ into  $- \sigma_y, - \sigma_x, - \sigma_z$. This restriction to the two selected nodes is applicable because  the eight nodes obtained from the node W1 by the $C_{2v}$ group operations $e, \sigma_v(xz), \sigma_v(yz), C_2$ and the time-inversion $T$ make coinciding contributions to $j_x^y$, and the remaining eight nodes also contribute equally.  For example, we demonstrate this for the nodes ${\bm k}_{W3} = \sigma_v(xz) {\bm k}_{W1}$  and ${\bm k}_{W4} = T {\bm k}_{W1} = -{\bm k}_{W1}$. In these valleys the components $d^{(Wi)}_{\alpha}(k_x,k_y,k_z)$ ($i=3, 4$) have the form 
\begin{eqnarray}
&&\mbox{}\hspace{1.3 cm}d^{(W3)}_x=- {\cal C}\hbar v_0 k_x - P_x(k_x,-k_y,k_z)\:, \nonumber\\
&&d^{(W3)}_y=- {\cal C}\hbar v_0 k_y + P_y(k_x,-k_y,k_z), d^{(W3)}_z= - {\cal C}\hbar v_0 k_z, \nonumber
\end{eqnarray}
and $d^{(W4)}_{\alpha}(k_x,k_y,k_z)=- d^{(W1)}_{\alpha}(-k_x,-k_y,-k_z)$.
One can check that the substitution of $d^{(Wi)}_{\alpha}({\bm k})$ into the sum~(\ref{C4v})  followed by the variable transformation $k_x, k_y, k_z$ to $k_x, -k_y, k_z$ or $-k_x, -k_y, -k_z$ completely restores the modulus $d({\bm k})$, and the product of functions to be summed over ${\bm k}$.

We take account for the additional terms $P_{\alpha}({\bm k})$ in the first order of the perturbation theory. There are several possible contributions to the photocurrent (\ref{general}) related to ${\bm P}({\bm k})$. They come from the corrections to (i)~the Berry curvature~(\ref{Berrygeneral}), (ii) the velocity ${\bm v}_{+-} ({\bm k})$ in Eq.~(\ref{general}), and (iii) the modulus $d({\bm k})$ in Eq.~(\ref{W+-}) as well as to the energy dependence of $\tau_p$ and $F({\bm k})$. Below we analyze these contributions, one after another.  To avoid cumbersome formulas we simplify the functions $P_x({\bm k}), P_y({\bm k})$  assuming that they depend on $k_x, k_y$ but are independent of $k_z$.

\subsubsection{Correction to the Berry curvature}

In the linear in $\bm P(\bm k)$ approximation one has for the Berry curvature in the valley W1
\[
\Omega^{(W1)}_{y}({\bm k}) = \frac{(\hbar v_0)^2}{2d^3} \left( {\cal C}\hbar v_0 k_y + P_y + k_y \frac{\partial P_x}{\partial k_x} - k_x \frac{\partial P_y}{\partial k_x}\right).
\]
In the W2 valley one has  $d^{(W2)}_x = -{\cal C}\hbar v_0 k_x - P_y(k_y, k_x)$, $d^{(W2)}_y =  -{\cal C}\hbar v_0 k_y - P_x(k_y, k_x)$, $d^{(W2)}_z =  - {\cal C}\hbar v_0 k_z$ and $d_0^{(W2)} = d_0(k_y,k_x,k_z)$. It follows from Eq.~(\ref{Berrygeneral}) that
\begin{eqnarray} \label{Omegay2}
&&\Omega^{(W2)}_{y}({\bm k}) = \frac{(\hbar v_0)^2}{2d^3} \left[ - {\cal C}\hbar v_0 k_y - P_x(k_y,k_x) \right.\\ &&\hspace{1 cm}\left. - k_y \frac{\partial P_y(k_y,k_x)}{\partial k_x} + k_x \frac{\partial P_x(k_y,k_x)}{\partial k_x}\right]. \nonumber
\end{eqnarray}
Substituting $\partial d/ \partial k_x = \hbar v_0 k_x /k$, $\Omega^{(W1)}_{y}({\bm k})$ or
$\Omega^{(W2)}_{y}({\bm k})$ into Eq.~(\ref{C4v}) and changing variables $k_x,k_y$ to $k_y,k_x$ in the sum for the W2 node contribution we obtain for the $N$-valley WSM
\begin{equation} \label{currentCURV}
j_x^y = \frac{4N \pi e^3 v_0^3 \tau_p  }{\hbar^2 \omega^3}  \left\vert {\bm E}\right\vert^2
\sum\limits_{\bm k} S_{\Omega} F({\bm k})\  \delta(2 \hbar v_0 k - \hbar \omega)\:,
\end{equation}
where
\begin{eqnarray}
S_{\Omega} &=& \frac{k_x}{k} \left( P_y({\bm k}) + k_y \frac{\partial P_x({\bm k})}{\partial k_x} - k_x \frac{\partial P_y({\bm k})}{\partial k_x} \right) \nonumber\\
&+& \frac{k_y}{k} \left( - P_x({\bm k}) - k_x \frac{\partial P_y({\bm k})}{\partial k_y} + k_y \frac{\partial P_x({\bm k})}{\partial k_y} \right)\:. \nonumber
\end{eqnarray}

\subsubsection{Correction to the velocity}

For the node W1, the corrections to the energy and the group velocity linear in ${\bm P}({\bm k})$ can be written as
\begin{equation} \label{Eapprox}
E_{\pm,{\bm k}} = d_0({\bm k}) \pm d({\bm k})\:,\: d({\bm k}) \approx \hbar v_0 k + {\cal C} \frac{ {\bm k} \cdot {\bm P}({\bm k})}{k}
\end{equation}
and
\begin{equation} \label{v+-1}
\delta v_{+-,x}({\bm k}) = \frac{2 {\cal C}}{\hbar k}  \left( P_x - \frac{k_x}{k} \frac{ {\bm k}\cdot {\bm P}}{k} +{\bm k} \cdot  \frac{\partial {\bm P}}{\partial k_x}\right)\:.
\end{equation}
The similar quantities for the node W2 are given by
\begin{eqnarray} 
&&d^{(W2)}({\bm k}) \approx \hbar v_0 k + {\cal C} \frac{ \tilde{\bm k} \cdot \tilde{\bm P}}{k}\:, \nonumber\\
&&\delta v^{(W2)}_{+-,x} = \frac{2 {\cal C}}{\hbar k}  \left(  \tilde{P}_y - \frac{k_x}{k} \frac{ \tilde{\bm k}\cdot  \tilde{\bm P}}{k} +\tilde{\bm k} \cdot  \frac{\partial  \tilde{\bm P}}{\partial k_x}\right), \label{v+-2}
\end{eqnarray}
 where $\tilde{\bm k}= \sigma_d{\bm k}= (k_y,k_x)$, $\tilde {\bm P} = {\bm P}(\sigma_d{\bm k})= {\bm P}(k_y,k_x)$. Substituting Eq.~(\ref{v+-1}) or (\ref{v+-2}) into Eq.~(\ref{C4v}) we obtain, similarly to the previous subsubsection, the equation which differs from Eq.~(\ref{currentCURV}) by the replacement $S_{\Omega} \to S_v$ where
\begin{equation}  \label{currentVELO}
S_v = \frac{k_y}{k} \left( P_x +{\bm k} \cdot  \frac{\partial {\bm P}}{\partial k_x} \right) - \frac{k_x}{k} \left( P_y  + {\bm k} \cdot  \frac{\partial {\bm P}}{\partial k_y} \right). \nonumber
\end{equation}
A sum of the contributions due to the corrections to the Berry curvature and velocity reduces to
\begin{eqnarray} 
\label{currentTOTAL}
&&j_x^y = \frac{4N \pi e^3 v_0^3 \tau_p }{\hbar^2 \omega^3}  \left\vert {\bm E}\right\vert^2
\sum\limits_{\bm k}  S \: F({\bm k}) \:  \delta(2 \hbar v_0 k - \hbar \omega)\:,\\
&&S = 2 \frac{k_x k_y}{k} \left( \frac{\partial P_x}{\partial k_x} - \frac{\partial P_y}{\partial k_y} \right) - \frac{k_x^2 - k_y^2}{k}  \left( \frac{\partial P_x}{\partial k_y} + \frac{\partial P_y}{\partial k_x}\right) \:.\nonumber
\end{eqnarray}

\subsubsection{Correction to the energy}

In this case the product $v_{+-}^x \Omega_y$ is proportional to $k_x k_y$ because of the linear dependence ${\bm d}({\bm k}) = {\cal C}\hbar v_0 {\bm k}$. This allows one to write the corresponding contribution to (\ref{C4v}) in the following general form
\[
{\cal C} \sum\limits_{\bm k} \frac{k_x k_y}{k} R[d({\bm k})] F({\bm k})\:,
\]
where $R$ is a function of the modulus $|{\bm d}({\bm k})|$. The  same sum related to the node W2 differs by the sign of the Berry curvature ${\cal C}$ and the variable change  in the  product 
\[
R[d({\bm k})] F({\bm k}) \to R[d(\sigma_d{\bm k})] F(\sigma_d{\bm k})\:.
\]
Changing the variable ${\bm k} \to \sigma_d {\bm k}$ in this sum and taking into account that the bilinear function $k_x k_y$ is invariant under the operation $\sigma_d$ we immediately find that the contributions from the nodes W1 and W2 differ in sign and cancel each other.  

It follows then that Eq.~(\ref{currentTOTAL}) gives the total photocurrent linear in ${\bm P}$ and this equation can be used as the working formula for calculations in particular cases of the wavevector-dependence of ${\bm P}$.

\subsection{Particular cases}
Taking $P_x = D_x k_x^n k_y^m$ and $P_y = D_y k_x^{n'} k_y^{m'}$  we obtain
\begin{eqnarray} \label{kS}
k S &=& D_x k_x^n k_y^{m-1}\ \left[ (2 n +m) k_y^2 - m k_x^2 \right] \\&+& D_y k_x^{n'-1} k_{y'}^{m'} \left[ - (2 m' +n') k_x^2 + n' k_y^2 \right]\:. \nonumber
\end{eqnarray}

\subsubsection{Quadratic nonlinearity}

For a combination of linear-${\bm k}$ and double-Weyl Hamiltonians with $P_x = D_{x1} \left( k_x^2 - k_y^2\right) + 2D_{x2}  k_x k_y$ and $P_y = D_{y1} \left( k_x^2 - k_y^2\right) + 2D_{y2}  k_x k_y$ one obtains from Eq.~(\ref{kS})
\begin{eqnarray} \label{Sfinal}
S &=& 2 k^2 \sin^3{\theta} \left[\left( D_{x1} - D_{y2} \right)\sin{3 \varphi} \right.\\ &&\mbox{} \hspace{5 mm}\left.- \left( D_{x2} + D_{y1} \right) \cos{3 \varphi} \right], \nonumber
\end{eqnarray}
where $\theta, \varphi$ are the polar angles of the wavevector ${\bm k}$. 
Therefore, the photocurrent becomes nonzero due to the angular harmonics of the third order contributing to the difference $F({\bm k})$ of the occupations of the initial and final states.
Then at $D_{x1} - D_{y2}=0$ we obtain from Eq.~\eqref{currentTOTAL}
\begin{equation}
j_x^y = -\left( D_{x2} + D_{y1} \right)\frac{Ne^3 \tau_p \omega}{8\pi\hbar^3 v_0^2}  \left\vert {\bm E}\right\vert^2 \left< {k_x(k_x^2-3k_y^2)\over k^3} F\left(\bm k\right)\right>,
\end{equation}
where the angular brackets mean averaging over directions of the vector $\bm k$ at a fixed absolute value ${k=\omega/(2v_0)}$.

To move further we consider the particular case of zero temperature and linear-${\bm k}$ dependence of the tilt ${d_0({\bm k}) = {\bm a} \cdot {\bm k} = a_x k_x + a_y k_y }$. Then for the negative chemical potential  $\mu = - |\mu|$ the function $F({\bm k})$ reduces to 
\begin{equation} \label{omegarange}
\Theta\left(- |\mu| + \hbar \omega/2 - {\bm a} \cdot {\bm k}\right) - \Theta\left(- |\mu| - \hbar \omega/2 -{\bm a} \cdot {\bm k}\right)\:,
\end{equation}  
where $\Theta(x)$ is the Heaviside step function.
For definiteness we set $a_y =0$ and $a_x > 0$, the general case of $a_x, a_y \neq 0$ is treated analogously. 

We introduce the coordinate system $(x'y'z')$ which is obtained from $(xyz)$ by a rotation around the $y$ axis by $90^\circ$: $x'=-z$, $y'=y$, $z'=x$.
Then the photocurrent is given by
\begin{align}
\label{j_int_t}
j_x^y = & -\left( D_{x2} + D_{y1} \right)\frac{Ne^3 \tau_p \omega}{8\pi\hbar^3 v_0^2}  \left\vert {\bm E}\right\vert^2 \\
& \times \left< {k_{z'}(k_{z'}^2-3k_{y'}^2)\over k^3} F\left({k_{z'}\over k}\right)\right>.
\nonumber
\end{align}
The average here can be presented as follows
\begin{equation}
\int\limits_{-1}^1 dt 
 {5t^3-3t\over 2} [\Theta\left(C- t \right) - \Theta\left(C'-t \right)],
\end{equation}
where
\begin{equation}
C={\hbar \omega -  2 |\mu| \over \hbar \omega b}, \qquad
C'={\hbar \omega +  2 |\mu| \over \hbar \omega b},
\end{equation}
and we use the dimensionless tilt parameter
\begin{equation}
\label{b_tilt_parameter}
b = {a_x \over \hbar v_0}.
\end{equation} 

In type-I WSMs where $b<1$, $C'>1$ for any $\omega$ and $\mu$, and the photocurrent is nonzero at $|C|<1$ only.
This gives the lower and higher edges for the optical absorption \cite{TypeIboundaries}
\begin{equation} \label{range1}
\frac{1}{1 + b} < \frac{\hbar \omega}{2 |\mu|} < \frac{1}{1 - b}\:.
\end{equation}
Integration yields
for type-I WSM
\begin{multline}
\label{j_typeI}
j_x^y = -\left( D_{x2} + D_{y1} \right)\frac{Ne^3 \tau_p |\mu|}{64\pi\hbar^4v_0^2}  \left\vert {\bm E}\right\vert^2 \\
\times {(1-C^2)(1 - 5C^2)\over 1-bC}\Theta(1-|C|).
\end{multline}

\begin{figure}[t]
\includegraphics[width=0.95\linewidth]{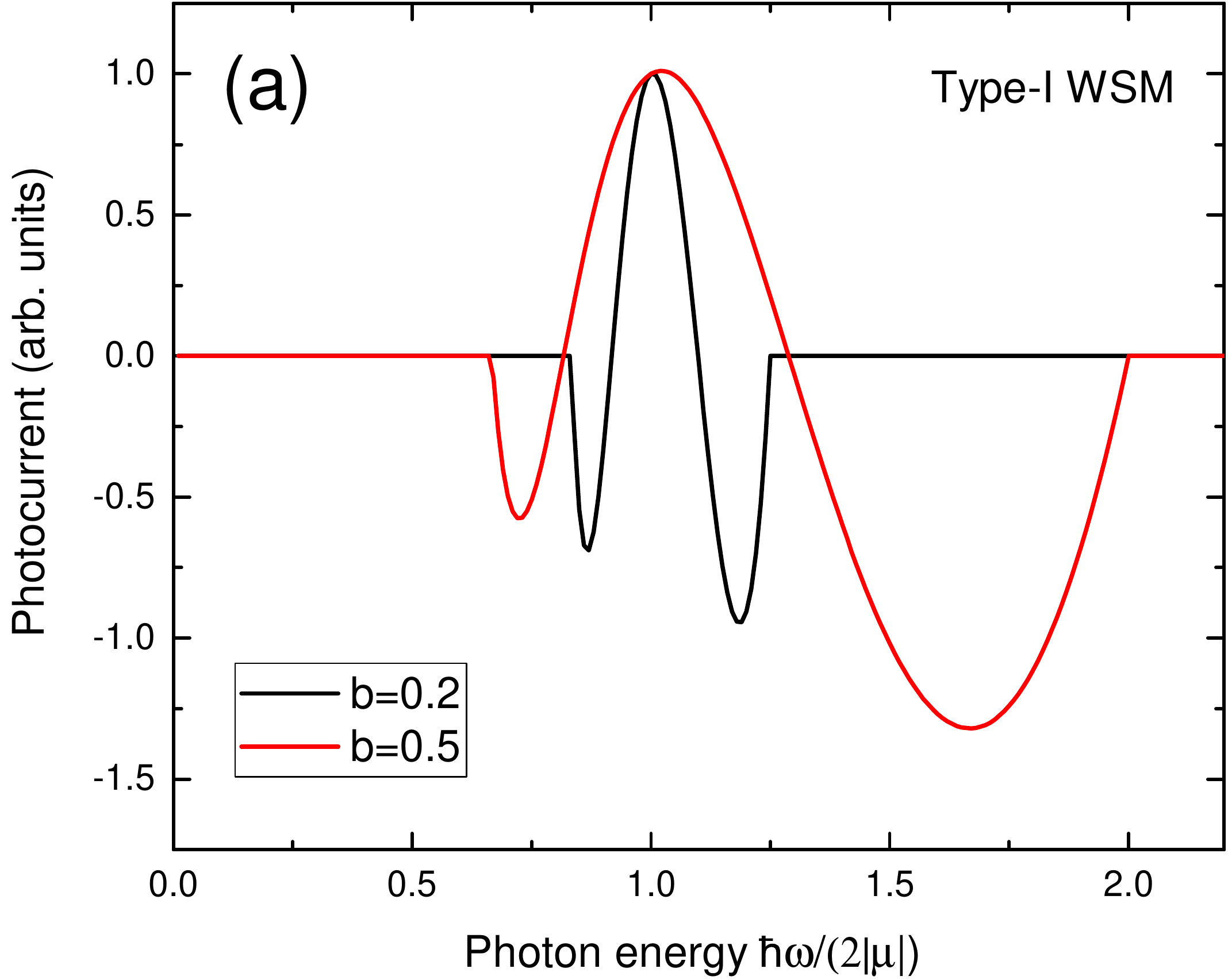}\\
\includegraphics[width=0.95\linewidth]{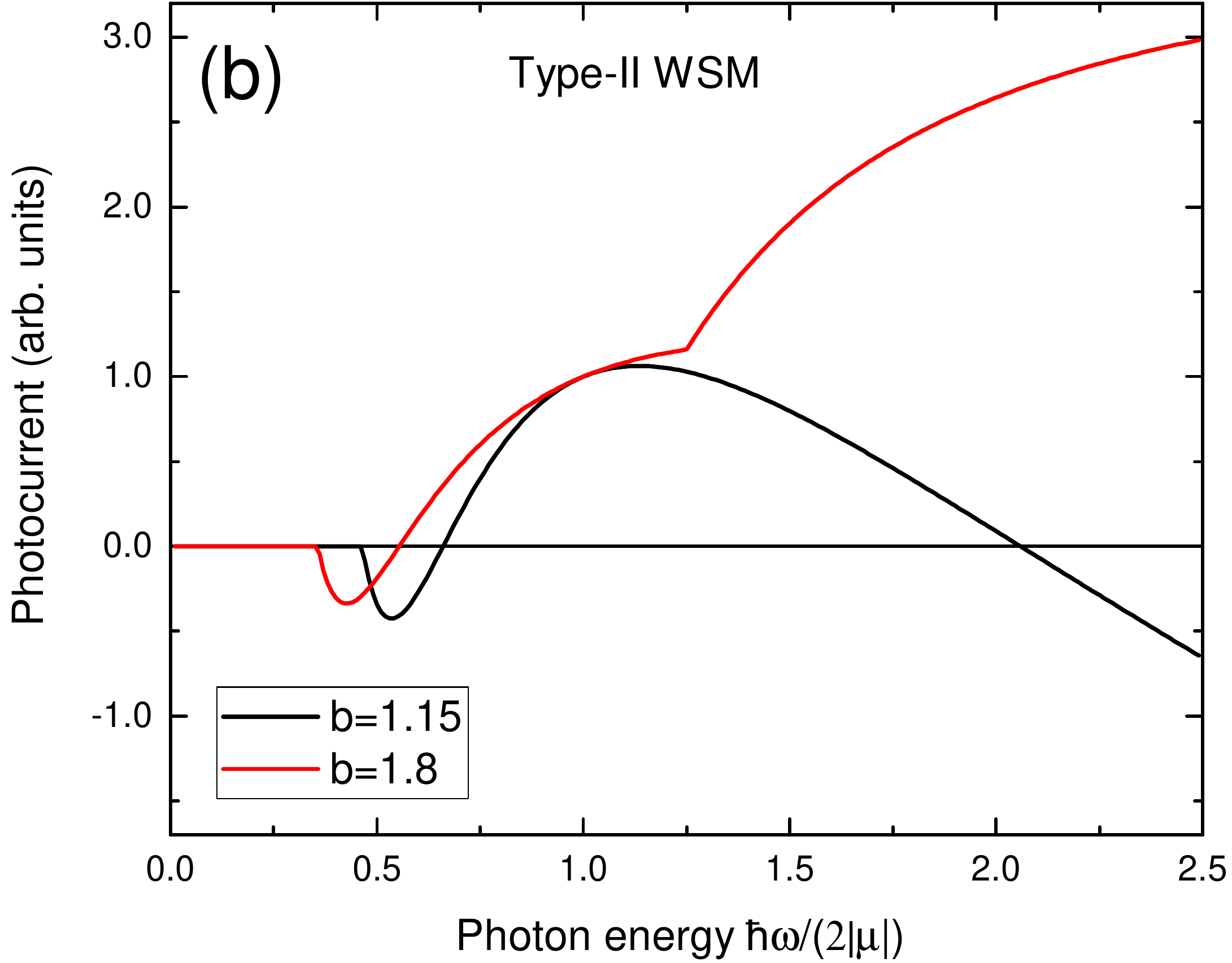}
\caption{Spectral dependence of the CPGE current calculated  in type-I (a) and type-II (b) WSMs for two values of the tilt parameter $b$ defined by Eq.~\eqref{b_tilt_parameter}.
}
\label{fig:spectra}
\end{figure}

In type-II WSMs where $b> 1$, the photocurrent is nonzero at $C>-1$. This means that
the low-frequency range is cut by \cite{AbsCircRad}
\begin{equation} \label{range2}
\frac{1}{b + 1} < \frac{\hbar \omega}{2 |\mu|}.
\end{equation}
For $C'>1$ i.e. for ${2|\mu|/\hbar\omega>b-1}$ the photocurrent in type-II WSM is also given by Eq.~\eqref{j_typeI}. For $C'<1$, i.e. for $2|\mu|/\hbar\omega<b-1$ it is equal to
\begin{align}
&j_x^y =\left( D_{x2} + D_{y1} \right)\frac{Ne^3 \tau_p |\mu|}{8\pi\hbar^4v_0^2}  \left\vert {\bm E}\right\vert^2 \\
&\times {1\over b^4}  \left\{ 5\left[1+\left({2|\mu|\over\hbar\omega}\right)^2 \right] -3b^2 \right\} \Theta\left({\hbar\omega \over 2|\mu| } -{1\over b - 1}\right). \nonumber
\end{align}

Excitation spectra for type-I and type-II WSM are shown in Fig.~\ref{fig:spectra}. The spectra are identical in the low-frequency part. At high frequencies the photocurrent is zero for type-I WSM while it raises with frequency in type-II WSM.

\subsubsection{Cubic nonlinearity}

In general the cubic contribution to $P_x(k_x,k_y)$, $P_y(k_x,k_y)$ can be presented as
\begin{equation} \label{Pcubic}
P_x = \sum\limits_{n=0}^3 D_{xn} k_x^nk_y^{3 - n}\:,\quad P_y = \sum\limits_{n=0}^3 D_{yn} k_x^{3 - n}k_y^n\:.
\end{equation}
Substitution of these expressions into Eq.~(\ref{Sfinal}) and averaging over the solid angle $4\pi$, hereafter indicated by a bar,  gives
\begin{equation} \label{Saverage}
\bar{S} = \frac{2 k^3}{15} \left[ 3\left( D_{x0} - D_{y0}\right) + D_{x2} - D_{y2} \right]\:.
\end{equation}
For illustration, in Ref.~\cite{JETP_Lett_2017}, we used the model with ${P_x = D k_y(k_x^2 + k_y^2)}$, $P_y = D k_x(k_x^2 + k_y^2)$. In the notation (\ref{Pcubic}) this corresponds to the choice $D_{x0} = D_{x2} = D_{y0} = D_{y2} = D$, $D_{x1} = D_{x3} =D_{y1} = D_{y3} =0$ in which case the average value  $\bar{S}$ vanishes and the circular photocurrent becomes nonzero only with allowance for the tilt. However, if  the combination of coefficients in square brackets of Eq.~(\ref{Saverage}) is nonzero then the photocurrent is induced even in the absence of tilt, particularly for $F({\bm k}) \equiv 1$ in Eq.~(\ref{currentTOTAL}).\\

\section{CPGE under indirect transitions}
\label{intraband}

Here we consider a WSM with a finite chemical potential $\mu>0$. At the photon energies $\hbar\omega<2\mu$, the light absorption is possible due to indirect transitions. We show that such indirect transitions are also accompanied by a photocurrent. For simplicity we assume a short-range scattering potential.

The photocurrent density is given by
\begin{equation}
j_z = e\sum_{\bm k \bm k'}W_{\bm k' \bm k} [v_z({\bm k'}) \tau_p(k')-v_z({\bm k}) \tau_p(k)].
\end{equation}
Here $W_{\bm k' \bm k}$ is the probability of the indirect optical absorption, and we introduced the momentum scattering time according to
\begin{equation}
	{1\over \tau_p(k)} = {2\pi \over \hbar} \sum_{\bm k'} |U^{cc}_{\bm k' \bm k} |^2 (1-\cos{\Phi})\delta[\hbar v_0 (k - k')]
\end{equation}
with $\Phi$ being the angle between ${\bm k}'$ and ${\bm k}$. For the short-range scattering potential the scattering matrix element $U^{cc}_{\bm k' \bm k}$ is equal to $U_0 \left<c\bm k'|c\bm k \right>$ where the Fourier image $U_0$ of the potential is a constant. After the summation we obtain
\begin{equation}
{1\over \tau_p(k)}={{\cal N} |U_0|^2 k^2 \over 3\pi\hbar^2 v_0} ,
\end{equation}
where ${\cal N}$ is the concentration of the static scatterers.

\begin{figure*}[t]
\includegraphics[width=0.85\linewidth]{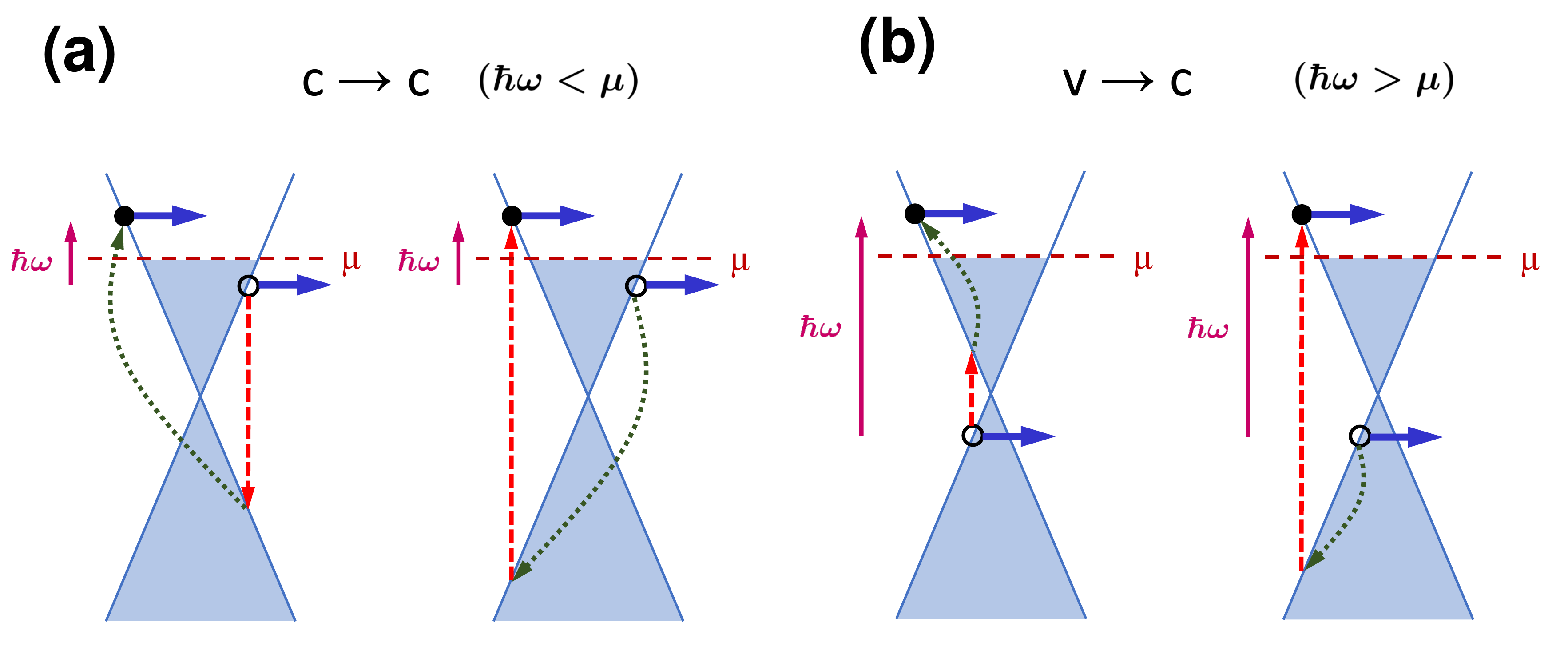}\caption{Quantum transitions resulting in the CPGE at ${\hbar\omega<2\mu}$. Dashed arrows indicate the direct optical transitions, and dotted arrows show the processes of impurity scattering. 
The photon energy $\hbar\omega$ shown by the length of solid vertical arrows is smaller~(a) and larger~(b) than $\mu$.
Only transitions with the initial state in the conduction band (a) are allowed at $\hbar\omega <\mu$. At $\hbar\omega >\mu$ the transitions with the initial state in the valence band (b) also contribute to the CPGE.}
\label{fig:Intraband}
\end{figure*}

\subsection{$c \to c$ processes}

The probability of the intraband optical absorption is described by the Fermi golden rule
\begin{equation}
W_{\bm k' \bm k} = {2\pi \over \hbar} |M_{\bm k' \bm k}|^2 (f_k-f_{k'})\delta(\hbar v_0 k'- \hbar v_0 k-\hbar\omega),
\end{equation}
where $f_k=f_0(\hbar v_0 k)$.
The electron-photon interaction operator is taken in the form
\begin{equation}
\label{e-phot}
V = {\cal C}v_0 {ie\over \omega} \bm\sigma \cdot \bm E \text{e}^{-i\omega t} + {\rm h.c.}
\end{equation}
Then the compound matrix element of the indirect optical transition via the valence band, Fig.~\ref{fig:Intraband}(a), reads
\begin{equation} \label{compound}
M_{c\bm k' c\bm k} = {U^{cv}_{\bm k' \bm k}V^{vc}_{\bm k} \over E_v(k)-E_c(k)-\hbar\omega}
+{V^{cv}_{\bm k'}U^{vc}_{\bm k' \bm k} \over E_v(k')-E_c(k)},
\end{equation}
with the interband scattering matrix elements being
\begin{equation}
U^{cv}_{\bm k' \bm k} = U_0 \left<c\bm k'|v\bm k \right>, \qquad U^{vc}_{\bm k' \bm k} = U_0 \left<v\bm k'|c\bm k \right>,
\end{equation}
and the interband optical matrix elements, for the $\sigma^+$ and $\sigma^-$ circular polarizations, being
\begin{eqnarray}
&&V^{cv}_{\bm k}(\sigma^+) = - V^{vc,*}_{\bm k}(\sigma^-) =  i{e |{\bm E}| v_0\over \omega} {{\cal C}\cos{\theta_{\bm k} + 1} \over \sqrt{2}}\text{e}^{i\varphi_{\bm k}},\nonumber\\
&&V^{vc}_{\bm k}(\sigma^+) = - V^{cv,*}_{\bm k}(\sigma^-) =i{e |{\bm E}| v_0\over \omega} {{\cal C}\cos{\theta_{\bm k} -  1} \over \sqrt{2}}\text{e}^{i\varphi_{\bm k}}.\nonumber
\end{eqnarray}
Note that the half difference of squared moduli of $V^{cv}_{\bm k}(\sigma^+)$ and $V^{cv}_{\bm k}(\sigma^-)$ yields Eq.~\eqref{MOmega}.

The energy conservation law and the electron-hole symmetry imply that
\begin{equation}
E_c(k') - E_c(k) = \hbar\omega, \quad E_v(k) = -E_c(k) = -\hbar v_0 k.
\end{equation}
Therefore the energy denominators in  Eq.~(\ref{compound}) coincide, and this equation  reduces to
\begin{equation}
 M_{c\bm k' c\bm k} = -{U^{cv}_{\bm k' \bm k}V^{vc}_{\bm k} + V^{cv}_{\bm k'}U^{vc}_{\bm k' \bm k}\over 2\hbar v_0 k + \hbar\omega}.
\end{equation}
As a result,  we obtain for the circular photocurrent density
\begin{eqnarray}
&&j_{z}^{c\to c}= {4e^3 v_0^3 \over 3\omega^2} P_\text{circ} {\cal C} |\bm E|^2   \\
&&\times \sum_{\bm k}  {f_0(\hbar v_0 k)-f_0(\hbar v_0 k+\hbar\omega)\over (2\hbar v_0 k + \hbar\omega)^2} \left[1+\left(1+{\omega\over v_0 k}\right)^2 \right]\:. \nonumber 
\end{eqnarray}

At  low temperatures integration yields
\begin{eqnarray}
\label{j_c_c}
j_{z}^{c\to c} &=& {4 \Gamma_0\over \pi \omega} P_\text{circ} {\cal C} |\bm E|^2 \\&& \times \left\{ 
\begin{array}{l}
\frac{1}{1 - (\hbar\omega/2\mu)^2}, ~~ \mbox{if}~~ \hbar\omega < \mu
\\ {2\mu(\hbar \omega+\mu)\over \hbar\omega(2\mu+\hbar\omega)}, ~~\mbox{if}~~  \mu < \hbar\omega < 2\mu
\end{array}
\right. . \nonumber
\end{eqnarray}

\subsection{Photocurrent caused by $v\to c$ transitions}

At $\hbar\omega >\mu$ but still $\hbar\omega<2\mu$ the transitions from the valence-band states also contribute to both light absorption and the CPGE, Fig.~\ref{fig:Intraband}(b). The probability rate and matrix element of the indirect transitions $v\to c \to c$ and $v\to v \to c$ are given by
\begin{equation}
W_{c\bm k' v\bm k} = {2\pi \over \hbar} (f_k^v-f_{k'})\delta(\hbar v_0 k'+ \hbar v_0 k-\hbar\omega)
|M_{c\bm k' v\bm k}|^2,
\end{equation}
\begin{equation}
M_{c\bm k' v\bm k} = {U^{cc}_{\bm k' \bm k}V^{cv}_{\bm k} + V^{cv}_{\bm k'}U^{vv}_{\bm k' \bm k}\over 2\hbar v_0 k - \hbar\omega},
\end{equation}
where $f_k^v=f_0(-\hbar v_0 k)$.
As a result, we obtain for the photocurrent density
\begin{equation}
j_{z}^{v\to c} = e{\cal C}v_0\sum_{\bm k \bm k'} [\tau_1(k') \cos{\theta_{\bm k'}} + \tau_1(k) \cos{\theta_{\bm k}}] W_{c\bm k' v\bm k},
\end{equation}
where we took into account that the initial electron state has the velocity $-{\cal C}v_0$.
Summation over $\bm k'$ yields
\begin{align}
 & j_{z}^{v\to c}=P_\text{circ}|\bm E|^2 \:{\cal C}  {8\Gamma_0\over \pi\omega} 
\\ & \times \int\limits_0^\infty dx {x^2+(1-x)^2\over (2x-1)^2} \bigl\{f_0(-x\hbar\omega)-f_0[(1-x)\hbar\omega]\bigr\}. \nonumber
\end{align}

At  low temperatures we have
\begin{equation}
j_{z}^{v\to c} = P_\text{circ}|\bm E|^2 \:{\cal C}  {4\Gamma_0\over \pi\omega} 
{2\mu(\hbar \omega-\mu)\over \hbar\omega(2\mu-\hbar\omega)}.
\end{equation}

\subsection{Total photocurrent at intraband absorption}

Summing up the contributions from both the $c\to c$ and $v \to c$ transitions, we obtain for  ${\hbar\omega > \mu}$
the dependence which is a continuation of the expression~\eqref{j_c_c} for $j_z^{c\to c}$ calculated at $\hbar\omega < \mu$.
Therefore, for the whole region $0 < \hbar\omega < 2\mu$ of intraband absorption, we have
\begin{equation}
j_z^\text{circ} = P_\text{circ}|\bm E|^2 \:{\cal C} {4\Gamma_0\over \pi \omega}{1\over 1 - (\hbar\omega/2\mu)^2}.
\end{equation}

We see that at  small photon energies $\hbar\omega \ll \mu$ (but still $\omega\tau \gg 1$) the CPGE current in the given Weyl node has a universal form
\begin{equation}
j_z^\text{circ}(\hbar\omega \ll \mu) = 
P_\text{circ}|\bm E|^2 \:{\cal C} {4\Gamma_0\over \pi \omega}
\end{equation}
determined by the fundamental constants and the frequency.

\section{Magnetogyrotropic photogalvanic effect}
\label{Sec_MPGE}

The MPGE electric current is induced under the unpolarized photoexcitation in the presence of a magnetic field and flows backwards with the field reversal. Here we consider strong magnetic fields where a current transverse to the magnetic field $\bm B$ is suppressed because of the quantized cyclotron motion. It follows then that, in WSMs of the $C_{2v}$ symmetry, the magnetic field can conveniently be applied along the $x'$ or $y'$ axis  introduced in Sect.~\ref{Linear_SOC} and making  the angle $45^\circ$ with the reflection planes $\sigma_v(xz), \sigma_v(yz)$. Phenomenologically the MPGE current density is described by~\cite{JETP_Lett_2017}
\begin{equation} \label{phenomquant}
j_{x'} = B_{x'} |\bm E|^2 \chi(B_{x'}^2), \quad
j_{y'} = -B_{y'} |\bm E|^2 \chi(B_{y'}^2),
\end{equation}
where $\chi$ is an even function of the magnetic field.

\subsection{Contribution of an individual Weyl node}
\label{MPGE_one_mode}
 
The magnetic field is included into the Weyl Hamiltonian by the Peierls substitution. 
In accordance with Eq.~(\ref{phenomquant})  the field ${\bm B}$ is assumed to be directed along the $x'$ axis. The energy spectrum consists of one chiral subband 
\begin{equation} \label{chiral}
E_0(p_{x'}) = -{\cal C}\text{sgn}(B_{x'}) \hbar v_0 k_{x'}
\end{equation}
and a series of the valence ($v$) and conduction ($c$) magnetic subbands enumerated by the positive integer ${n= 1,2\dots}$ and having the dispersion relations ${E_{n}^{(c,v)}(k_{x'}) = \pm E_n(k_{x'})}$, where
\begin{equation}
E_n(k_{x'}) = \hbar\sqrt{(v_0k_{x'})^2+n\omega_c^2}
\end{equation}
and 
\begin{equation}
\omega_c = {v_0\sqrt{2}\over l_B}, \quad l_B=\sqrt{\hbar c\over |eB_{x'}|}.
\end{equation}
For $B_{x'}>0$,  the wavefunctions are given by
\begin{equation} \label{wavefunc}
	\Psi_0 = 	\left[ 
	\begin{array}{c}
		0\\
		\Phi_0
	\end{array}
	\right], \qquad
		\Psi_{n>0}^{(c,v)} = 
		\left[ 
	\begin{array}{c}
		a_{n,\pm}\Phi_{n-1}\\
		\pm {\cal C} a_{n,\mp}\Phi_n
	\end{array}
	\right],
\end{equation}
while for the opposite direction, $B_{x'}<0$, they 
are obtained from the above functions by the transformation
\begin{equation}
\Psi_{n}^{(c,v)}(B_{x'}<0) = -i{\cal C}\sigma_2 \Psi_{n}^{(v,c)}(B_{x'}>0).
\end{equation}
Here $\sigma_2$ is the second Pauli matrix,
\begin{equation}
\Phi_n = \text{e}^{i(k_{x'}x'+k_{y'}y')}\phi_n\left(z-{\hbar k_{y'} c\over eB_{x'}} \right),
\end{equation}
$\phi_n$ are the Landau-level oscillator functions, and the coefficients $a_{n,\pm}= \sqrt{(1\pm {\cal C} \hbar v_0 k_{x'}/ E_n)/2}$.

By using the wavefunctions (\ref{wavefunc})  and the perturbation (\ref{e-phot}) one can find the squared matrix element $| M_{c,n' \leftarrow v,n} |^2$ for the direct optical transitions of electrons between the $n$th magnetic subband in the valence band and the $n'$th conduction subband. For the linear polarization, $\bm E_\perp \perp \bm B$, the selection rules are $n' - n = \pm 1$.  The direct calculation gives
\begin{align}
\label{M_sq_inter}
&	\left| M_{c,n+1 \leftarrow v,n} (k_{x'})\right|^2 = \left({ev_0|\bm E_\perp|\over 2\omega}\right)^2 \\
& \times \left[ 1+{(\hbar v_0 k_{x'})^2\over E_nE_{n+1}} + {\cal C}\hbar v_0\text{sgn}(B_{x'})k_{x'} \left({1\over E_n} + {1\over E_{n+1} } \right)\right],\nonumber
\end{align}
\begin{equation}
\label{eh_time_inv}
\left| M_{c,n \leftarrow v,n+1}(k_{x'}) \right|^2 = \left| M_{c,n+1 \leftarrow v,n} (-k_{x'})\right|^2,
\end{equation}
for $n, n' \neq 0$, and 
\begin{eqnarray} \label{10-01}
&&\left| M_{c1 \leftarrow 0}(k_{x'}) \right|^2  = \left| M_{0 \leftarrow v1}(-k_{x'}) \right|^2 \\
&&\hspace{1 cm}= {(ev_0|\bm E_\perp|)^2\over 2\omega^2} 
\left[ 1+ { {\cal C}\hbar v_0\text{sgn}(B_{x'})k_{x'} \over E_1} \right]\nonumber
\end{eqnarray}
for $n= 0$ either $n'=0$.
Here the even in $k_{x'}$ terms are responsible for light absorption~\cite{magn_opt_cond}.

\begin{figure}[t]
\includegraphics[width=0.8\linewidth]{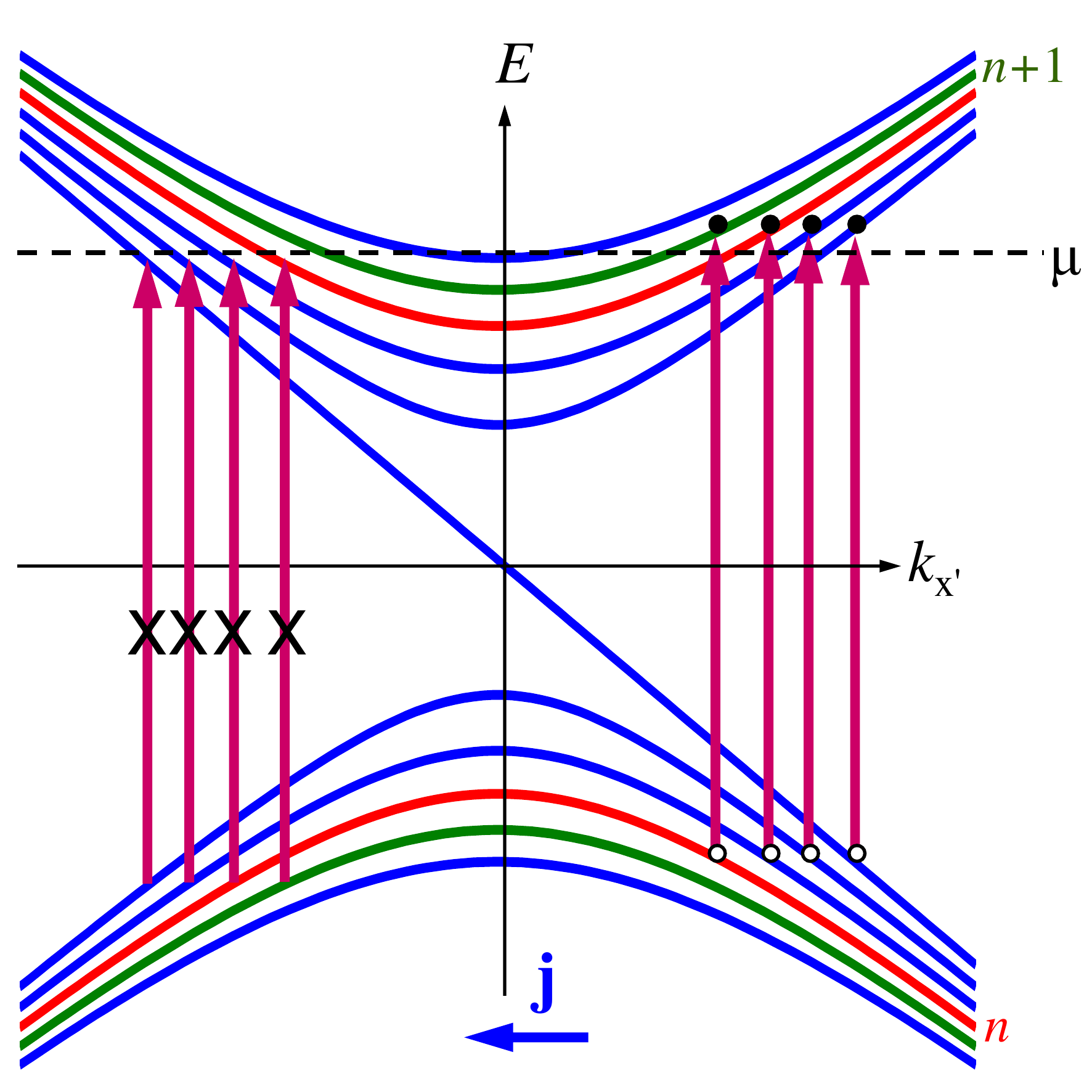}
\caption{Scheme of direct optical transitions between the one-dimensional subbands in a quantized magnetic field in the case ${\cal C}B_{x'}>0$. The transitions  $v,n \to c,n+1$ result in the MPGE current since the $v,n+1 \to c,n$ transitions are blocked (crosses) by the Pauli principle. For $n\neq0$ each transition occurs at two points $k_{x'}$ of opposite signs. The arrows illustrate the optical transitions, their starting points (open circles) are chosen at the side with the more probable transition rate.}
\label{fig:MPGE_1}
\end{figure}

Equations (\ref{M_sq_inter}) and (\ref{eh_time_inv}) allow one to explain the origin of a photocurrent under the $v,n \rightarrow c, n+1$ transitions with $n> 0$. 
Due to the terms odd both in $k_{x'}$ and $B_{x'}$, the probability rates are asymmetric with a predominance of states with ${\cal C}{\text{sgn}(B_{x'})k_{x'}>0}$ for the transitions {$v,n \rightarrow c, n+1$} and ${\cal C}{\text{sgn}(B_{x'})k_{x'}<0}$ for the transitions {$v, n+ 1 \to c,n$}. In these two kinds of transitions the electron energies of the initial (or final) states are different, Fig.~\ref{fig:MPGE_1}, and, therefore, their equilibrium occupation can be also different. This gives rise to the MPGE current in the polarization $\bm E \perp \bm B$. 
For the light polarized along the magnetic field, the selection rules read $n'=n$, the squared matrix elements are even in $k_{x'}$ and the electron photoexcitation is not accompanied by the current generation.

It is instructive to divide the relevant light spectral area into three ranges as presented below.

\subsubsection{Range 1: $\omega/\omega_c>\sqrt{2}+1$}

The density of the interband photocurrent is given by 
\begin{align}
\label{MPGE}
j_{x'}= {e \over \hbar l_B^2}  \sum_{n,k_{x'}} & (v^{x'}_{n+1}\tau_{n+1} +v^{x'}_n \tau_n)  \left| M_{c,n+1 \leftarrow v,n} \right|^2 \\ 
& \times F \: \delta \left[E_{n+1}(k_{x'}) + E_n(k_{x'}) -\hbar\omega\right]. \nonumber
\end{align}
Here $\tau_n$ is the momentum relaxation time in the $n$th magnetic subband, $v^{x'}_n$ is the velocity $\hbar v_0^2 k_{x'} / E_n$,
\begin{equation}
\label{F_MPGE}
F= (f^v_{n}-f^c_{n+1}) -(f^v_{n+1}-f^c_{n}),
\end{equation}
$f_n^{c,v}$ are the equilibrium occupations of the $n$th subbands in the conduction and valence bands. While deriving Eq.~(\ref{MPGE}) we took into account {that} the  odd parts of the squared matrix elements (\ref{eh_time_inv}) are opposite in sign but equal in magnitude.

The interband transitions {$v,n \to c,n+1$}, Fig.~\ref{fig:MPGE_1},
contribute to the photocurrent with $n$ satisfying the cut-off frequency relation
\begin{equation}
 \hbar\omega_c(\sqrt{n}+\sqrt{n+1}) \leq \hbar\omega, \hspace{1 mm}\mbox{or} \hspace{1 mm}  n\leq \nu \equiv  {(\omega^2-\omega_c^2)^2 \over 4\omega^2\omega_c^2}.
\end{equation}
For $n~\neq~0$ the direct transitions occur at the points $k_{x'} = \pm k_\omega^{(n)}$.
The quasi-momentum $\hbar k_\omega^{(n)}$ as well as the initial and final energies, $-E_n(\omega)$ and $E_{n+1}(\omega)$,  are found from the energy conservation ${E_n (k_\omega^{(n)}) + E_{n+1}(k_\omega^{(n)}) = \hbar \omega}$ as follows
\begin{equation}
\label{En}
\hbar k_\omega^{(n)} = 
{E_n (\omega)\over v_0}\sqrt{1-{n\over \nu}}, 
\: E_{n/n+1}(\omega) = \hbar {\omega^2\mp\omega_c^2\over 2\omega}.
\end{equation}
For $n=0$, i.e. for the process {$0 \to c1$}, the transition occurs in the point $k_{x'} = k_\omega^{(0)}$ if ${\cal C} B_{x'} > 0$ and the point $k_{x'} = - k_\omega^{(0)}$ if ${\cal C} B_{x'} < 0$. In the former case Eq.~(\ref{En}) is also valid if $E_n$ is replaced by $- E_0\left(k_\omega^{(0)}\right)$, where $E_0\left(k_\omega^{(0)}\right)$ is the initial energy in the chiral subband, Eq.~(\ref{chiral}). We draw attention to the fact that the energies of electrons involved in the transitions {$v,n \to c,n+1$} are independent of $n$ {and shifted} relative to those in the {$v,n+1 \rightarrow c,n$} transitions by $\hbar \omega^2_c/\omega$.
 
We start the analysis from zero temperature, the effect of temperature is considered in Sec.~\ref{T}. 
In Range 1 the contributions of the transitions $v,n \to c, n+1$ and $v, n+1 \to c, n$ to the photocurrent do not compensate each other if the latter transition is suppressed by the Pauli principle, Fig.~\ref{fig:MPGE_1}. This takes place at values of the chemical potential $\mu>\hbar\omega_c$ where $f^c_{n+1}=0$ but $f^c_{n}=1$, i.e. at 
\begin{equation}
\label{B_range}
E_n(\omega) < \mu < E_{n+1}(\omega), \hspace{1 mm}\mbox{or} \hspace{1 mm}  |2\mu -\hbar\omega| < {\hbar\omega_c^2\over \omega}, 
\end{equation}
or, equivalently, in the frequency range ${\Omega^-(\mu)<\omega < \Omega^+(\mu)}$, where
\begin{equation}
\label{range_omega}
\hbar\Omega^\pm(\mu) = \mu + \sqrt{\mu^2\pm(\hbar\omega_c)^2}.
\end{equation}
In fact, the lower frequency edge is determined by the largest value between $(1 + \sqrt{2}) \hbar \omega_c$ and $\mu + \sqrt{\mu^2-(\hbar \omega_c)^2}$. The crossover between these two values  occurs at the point
$\mu = \sqrt{2} \hbar \omega_c$.

Assuming the momentum relaxation times in all magnetic subbands to coincide and be equal to $\tau_p$, we obtain
\begin{equation}
\label{j_inter}
j_{x'} ={\cal C}\text{sgn}(B_{x'}) \Gamma_0 \tau_p |\bm E|^2 
{2\omega_c^2 \over \omega^2 + \omega_c^2}  \sum_{0\leq n<\nu}  \sqrt{1- {n\over \nu}},
\end{equation}
where $\Gamma_0$ is introduced in Eq.~(\ref{Gamma00}) and we replaced $|{\bm E}_{\perp}|^2$ by $(2/3)|\bm E|^2$.

In the weaker (but still quantized) fields such as $\omega_c \ll \omega\approx 2\mu/\hbar$, we have $\nu \approx (\omega/2\omega_c)^2 \gg 1$ and, replacing the summation by integration, obtain a universal result
\begin{equation}
\label{MPGE_universal}
	j_{x'} (\omega_c \ll \omega)= {\cal C}\text{sgn}(B_{x'}) {\Gamma_0 \over 3}\tau_p |\bm E|^2. 
\end{equation}

\subsubsection{Range 2: $1<\omega/\omega_c<\sqrt{2}+1$} 

\begin{figure}[t]
\includegraphics[width=0.8\linewidth]{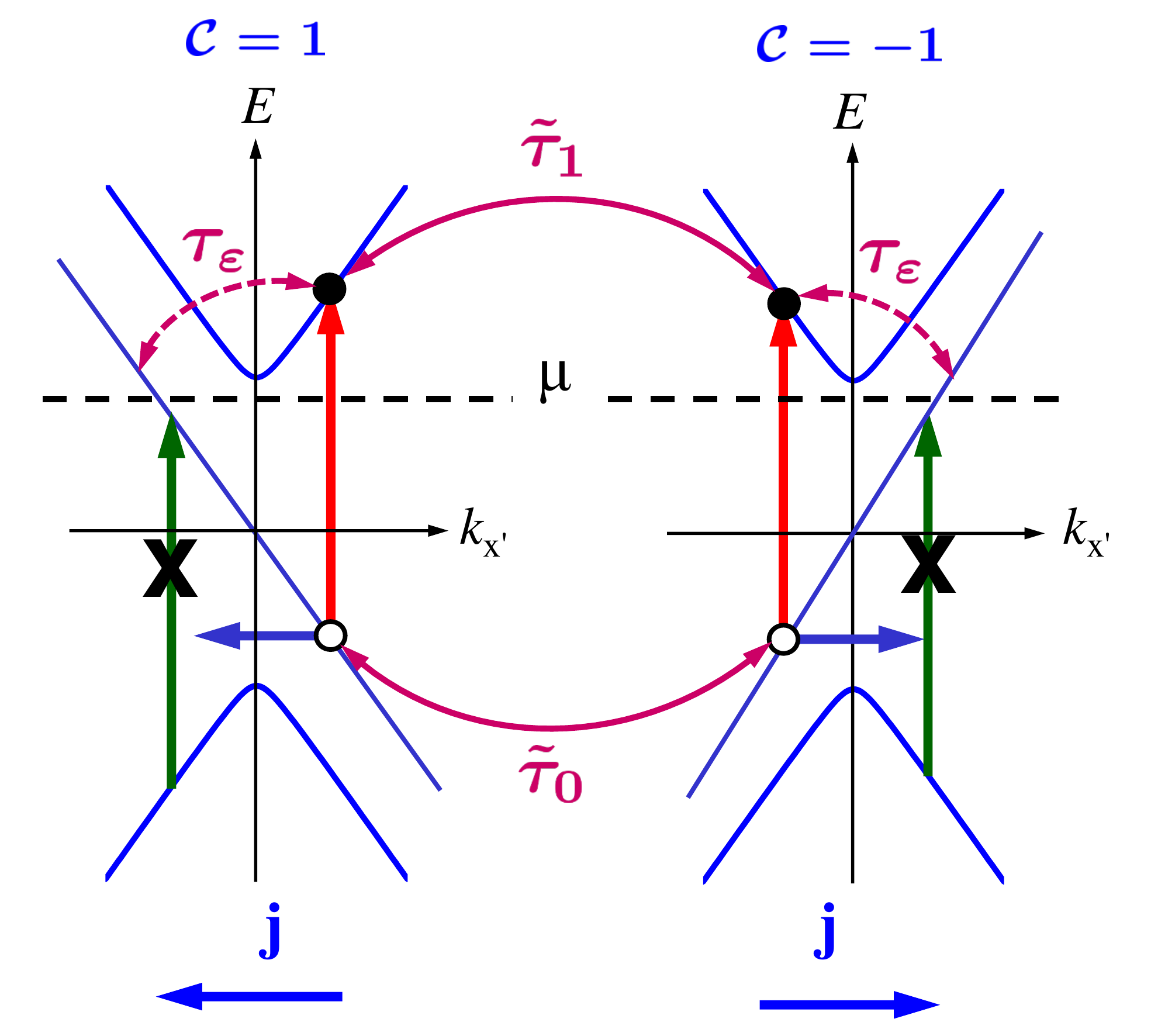}
\caption{Schematics of the magneto-gyrotropic photogalvanic effect and the related scattering times controlling the photocurrent formation. The current is governed by a combination of three relaxation times, namely, the intranode energy relaxation time $\tau_\varepsilon$ and the internode elastic scattering times between the chiral subbands ($\tilde{\tau}_0$) and between the first excited subbands ($\tilde{\tau}_1$), see Eq.~\eqref{tau_MPGE}. The position of chemical potential $\mu$ prevents the transitions from the valence subband $v1$ to the chiral subband with the linear dispersion.}
\label{fig:rel_times}
\end{figure}

The elastic scattering of electrons (or holes) in the chiral subband (\ref{chiral}) with the energies within the interval $(- \hbar \omega_c, \hbar \omega_c)$ radically differs from that in all other subbands. For energies outside this interval,
the backward scattering of the quasi-momentum within the same Weyl node is the main mechanism of the free-carrier momentum relaxation resulting in a short intravalley relaxation time $\tau_p$ comparable to the zero-field one. 
However, the energies between $- \hbar \omega_c$ and $\hbar \omega_c$
are available only for the states in the chiral subband which has a linear dispersion and, therefore, forbids an intravalley backscattering. In this case the momentum relaxation occurs due to scattering to the chiral subbands in other Weyl points, Fig.~\ref{fig:rel_times}, with a characteristic time $\tilde{\tau}_0$ obviously by far exceeding $\tau_p$.

A photohole excited in the transition $0 \rightarrow c1$ in Range~2 has an energy $|E_0|$ below $ \hbar \omega_c$. The consideration of steady-state kinetics yields the following effective relaxation time which controls the photocurrent~\cite{JETP_Lett_2017}:
\begin{equation}
\label{tau_MPGE}
		\tilde{\tau} = {\tilde{\tau}_0\tau_\varepsilon \over \tilde{\tau}_1} \gg \tau_p.
\end{equation}
Here $\tilde{\tau}_1$ is the time of elastic scattering of carriers between the 1st subbands in different monopoles, and $\tau_\varepsilon$ is the energy relaxation time describing phonon-involved transitions from the excited states in the 1st conduction subband to the 0th chiral subband, Fig.~\ref{fig:rel_times}. These scattering processes in WSMs and their effect on CPGE are discussed in Ref.~\cite{CPGE_scattering}. To summarize, the contribution to the photocurrent from the photoholes in the chiral subband dominates and one has for the photocurrent density
\begin{align}
\label{j01}
j_{x'} = -{e  \tilde{\tau} \over \hbar l_B^2}\sum_{k_{x'}} & v^{x'}_0(k_{x'}) \left\vert M_{c1 \leftarrow 0}(k_{x'}) \right\vert^2 \\ & \times \delta \left[E_1(k_{x'}) - E_0(k_{x'}) -\hbar\omega\right]\:. \nonumber
\end{align}
Here the chemical potential lies between the Weyl-point energy and the energy $\sqrt{2} \hbar\omega_c$, and the temperature is set to zero. For $\hbar \omega_c > \mu > 0$, the frequency range narrows to $(\hbar \omega_c, \mu + \sqrt{\mu^2 + (\hbar \omega_c)^2})$, while for $\sqrt{2} \hbar \omega_c > \mu > \hbar \omega_c $ the photocurrent is generated in the range $[\mu + \sqrt{\mu^2  - (\hbar \omega_c)^2}, \hbar \omega_c(1 + \sqrt{2})]$.

The  squared matrix element $\left| M_{c1 \leftarrow 0} \right|^2$ is even in magnetic field because, under the inversion of $B_{x'}$, the slope of dispersion $E_0(p_{x'})$ changes from negative to positive, and the direct transition $0 \to c1$ is replaced by the transition $v1 \to 0$ that takes place in the inverted value of $k_{x'}$. However the velocity in the chiral subband $v^{x'}_0 = - {\cal C}v_0\text{sgn}(B_{x'})$ is odd in $B_{x'}$, and we obtain~\cite{JETP_Lett_2017}:
\begin{equation}
\label{chircurrent}
j_{x'} = B_{x'}|\bm E|^2 \: {\cal C} \Gamma_0\tilde{\tau} {2v_0^2e\over \hbar c \omega^2}\:.
\end{equation}
For $B_{x'}=1$~T, $v_0=c/300$, and $\omega/(2\pi)=1$~THz a value of $j_{x'}$ is estimated as $j_0 =  80 \: \Gamma_0\tilde{\tau}|\bm E|^2$.

\subsubsection{Range 3: $\omega<\omega_c$. Direct intraband transitions}

If the chemical potential lies in the energy interval $(0,\hbar \omega_c)$ then the main contribution to the photocurrent comes from the $0 \to c1$ transition under the excitation in the window 
\begin{equation} \label{chiral3}
\sqrt{\mu^2 + (\hbar\omega_c)^2} - \mu < \hbar \omega < \hbar\omega_c 
\end{equation}
and is described by Eq.~(\ref{chircurrent}). Thus, we turn to the samples with chemical potential located above $\hbar \omega_c$, i.e. above the bottom of the 1st conduction subband.

At zero temperature the direct optical transitions $c,n \rightarrow c,n+1$ ($n > 0$) between the conduction subbands occur if the chemical potential lies above the bottom of the $c1$ subband and $n < \nu$~\cite{magn_opt_cond}. The intraband absorption is accompanied by a photocurrent generation because in this case the probability rate as well contains a part odd in $k_{x'}$. The corresponding photocurrent has the form
\begin{eqnarray}
j_\text{intra} &=& {e\tau_p\over \hbar l_B^2} \sum_{k_{x'}, n\leq \nu} \left| M^\text{odd}_{c,n+1 \leftarrow c,n}(k_{x'}) \right|^2 \\
&&\times (v^{x'}_{n+1}-v^{x'}_{n}) \delta[E_{n+1}(k_{x'})-E_{n}(k_{x'})-\hbar\omega]. \nonumber
\end{eqnarray}
Here the transition matrix element is given by Eq.~\eqref{M_sq_inter} where $E_{n+1}$ should be replaced by $-E_{n+1}$.
The electron energies of the initial and final states are equal to
\begin{equation}
E_n/E_{n+1}= \hbar {\omega_c^2 \mp \omega^2\over 2\omega}\:.
\end{equation}
The conditions $E_{n+1} - E_n = \hbar \omega$, $E_n < \mu < E_{n+1}$ and $n> 0$ restrict the frequency range to
\begin{equation}
\label{range_omega2}
\hbar\omega \in \left(\sqrt{\mu^2 + (\hbar\omega_c)^2} - \mu, \mu - \sqrt{\mu^2-(\hbar\omega_c)^2} \right). 
\end{equation}
Analytically the photocurrent $j_{x'}$ is described by the same equation Eq.~\eqref{j_inter} derived 
for the interband excitation bearing in mind that now $\omega < \omega_c$.

\subsubsection{Temperature dependence}
\label{T}

The discontinuities in the dependence of the MPGE current on the magnetic field are smeared by temperature $T$. At finite $T$, the populations of the initial and final states take values other than just 0 and 1. As a result, the  ``complementary'' transitions ${v,n+1 \rightarrow c,n}$ forbidden in Fig.~\ref{fig:MPGE_1} by the Pauli exclusion principle become possible.
Using Eq.~\eqref{En} for the initial and final energies  we obtain for the populations
\begin{eqnarray}
&&f^{c,v}_n=f_0\left(\pm{\hbar(\omega^2-\omega_c^2)\over2\omega}\right), \: \\
&&f^{c,v}_{n+1}=f_0\left(\pm{\hbar(\omega^2+\omega_c^2)\over2\omega}\right).\nonumber
\end{eqnarray}
Substitution of these functions into Eqs.~\eqref{MPGE} and~\eqref{F_MPGE} results in the following temperature dependence of the photocurrent for the interband transitions 
\begin{equation}
\label{nonzero_T}
j_{x'} = j_{x'}(T=0) {2 \sinh{\left({\hbar\omega_c\over 2T}\right)}\sinh{\left({\hbar\omega\over 2T}\right)} \sinh{\left({\mu\over T}\right)} \over Z},
\end{equation} 
where 
\begin{eqnarray}
Z &=& \left[\cosh{\left({\mu\over T}\right)} + \cosh{\left({\hbar\omega_c\over 2T}\right)}\cosh{\left({\hbar\omega\over 2T}\right)}  \right]^2 \nonumber \\&& \hspace{8 mm} - \left[ {\sinh{\left({\hbar\omega_c\over 2T}\right)}} \sinh{\left({\hbar\omega\over 2T}\right)}  \right]^2. \nonumber
\end{eqnarray}

\subsection{Allowance for tilt}

The derived expressions for the MPGE current are different in sign for monopoles of opposite chirality. Here we demonstrate that account for the tilt terms in the Hamiltonian gives rise to the net MPGE current in gyrotropic WSMs.

In the linear-in-${\bm k}$ approximation the tilt term in Eq.~(\ref{Hamilt3}) can be written as $d_0({\bm k}) = {\bm a} \cdot {\bm k}$ and is  characterized by the vector $\bm a$ which describes the magnitude of a spin-independent correction to the Hamiltonian. Let us use, instead of $x',y',z$, the axes $x',y'', z''$ with 
$y''$ being the axis along the component ${\bm a}_{\perp}$ of the tilt vector $\bm a$  perpendicular to $\bm B$. Then  the Hamiltonian~(\ref{lineartilt}) takes the form
\begin{equation}
	{\cal H}={\cal C}\hbar v_0\bm \sigma \cdot \bm k + a_{x'} k_{x'} + a_{\perp} k_{y''}.
\end{equation}

With account for tilt, the energy dispersion in the magnetic subbands transforms to
\begin{align}
&E_0 = [-{\cal C}\hbar\text{sgn}(B_{x'})\tilde{v}_0 + a_{x'}] k_{x'}, \\ 
&E_{n}^{(c,v)} = a_{x'} k_{x'} \pm \tilde{E}_n. \nonumber
\end{align}
Here $\tilde{E}_n = \hbar\sqrt{\left(\tilde{v}_0 k_{x'}\right)^2 + n\tilde{\omega}_c^2}$,
the tilde marks the renormalized Fermi velocity, cyclotron frequency and magnetic length
\begin{equation}
 \tilde{v}_0 = {v_0\over \gamma},\quad
   \tilde{\omega}_c = {\tilde{v}_0 \sqrt{2}\over \tilde{l}_B}=\omega_c\gamma^{3/2}, \quad \tilde{l}_B= l_B\sqrt{\gamma},
\end{equation}
where 
\begin{equation}
\gamma = {1\over \sqrt{1-\beta^2}}, \qquad \beta = {a_{\perp} \over \hbar v_0}.
\end{equation}
At $B_{x'} > 0$, the wavefunctions in the conduction and valence magnetic subbands are given by~\cite{PRL_2016_1,PRL_2016_2,PRL_2016_3}
\begin{equation}
	\Psi_0 = {\text{e}^{i\alpha}\over \sqrt{\gamma}}	\text{e}^{-\theta\sigma_1/2}\left[ 
	\begin{array}{c}
		0\\
		\phi_0(z''-\zeta_0)
	\end{array}
	\right],
\end{equation}
\begin{equation}
		\Psi_{n>0}^{(c,v)} = 
		{\text{e}^{i\alpha}\over \sqrt{\gamma}}	\text{e}^{-\theta\sigma_1/2}\left[ 
	\begin{array}{c}
		\tilde{a}_{n,\pm}\phi_{n-1}\left(z''-\zeta_{n,\pm}\right)\\
		\pm {\cal C}  \tilde{a}_{n,\mp}\phi_n\left(z''-\zeta_{n,\pm}\right)
	\end{array}
	\right],
\end{equation}
where $\sigma_1$ is the first Pauli matrix, $\alpha=k_{x'}x'+k_{y''}y''$, $\tilde{a}_{n,\pm}= \sqrt{\left(1\pm {\cal C}\hbar{\tilde{v}_0 k_{x'}/ \tilde{E}_n}\right)/2}$, $\tanh{\theta} = \beta$, and the centers of the cyclotron orbits depend on the energies due to the tilt 
\begin{eqnarray}
&&\zeta_0 = {c\hbar\over eB_{x'}}\left(k_{y''} - \beta\gamma{\cal C}k_{x'}  \right),\\
&&\zeta_{n,\pm}={c\hbar\over eB_{x'}}\left(k_{y''} \pm {\beta\gamma\tilde{E}_n\over \hbar \tilde{v}_0}   \right). \nonumber
\end{eqnarray}

We calculate the photocurrent due to the $0 \to c1$ transitions assuming that the chemical potential lies below the bottom of the $c1$ subband: $0<\mu<\hbar\omega_c$, Fig.~\ref{fig:j_10_tilt}.
At ${\omega/\tilde{\omega}_c<\sqrt{2}+1}$, the photocurrent is given by Eq.~\eqref{j01} where the substitutions $l_B \to \tilde{l}_B$, $E_{0,1} \to \tilde{E}_{0,1}$ are made and the velocity in the 0th subband is 
$$v^{x'}_0= - {\cal C}\text{sgn}(B_{x'})\tilde{v}_0 + {a_{x'} \over \hbar}.$$
The electron-photon interaction operator can  still be taken in the form of Eq.~\eqref{e-phot} because
the tilt term does not result in interband transitions.
Using the relations
\begin{equation}
\left< \phi_0(z''- \zeta_0) \big| \phi_1\left(z''-\zeta_{1,+}\right) \right> = \sqrt{u}\text{e}^{-u/2}, 
\end{equation}
\begin{equation}
\left< \phi_0(z''-\zeta_0) \big| \phi_0\left(z''-\zeta_{1,+}\right) \right> = \text{e}^{-u/2},
\end{equation}
where $u=\left[(\zeta_{1,+} - \zeta_0)/(\tilde{l}_B\sqrt{2})\right]^2$,
we obtain that the matrix element of the direct optical transition is given by
\begin{align}
& M_{c1 \leftarrow 0} = {\cal C}\tilde{v}_0 {ie\over \omega}\text{e}^{-u/2} \\
& \times \left[ 
\gamma E_{y''} (\tilde{a}_{1,+} - \beta \sqrt{u}\tilde{a}_{1,-}) - i\tilde{a}_{1,+}E_{z''} - \sqrt{u}\tilde{a}_{1,-}E_{x'} \right]. \nonumber
\end{align}
Energy conservation yields $u = (\beta \omega/\tilde{\omega}_c)^2$, and
we 
obtain the MPGE current with allowance for tilt in the form
\begin{eqnarray}
\label{j_tilt}
j_{x'} &=& \left[{\cal C} \text{sgn}(B_{x'})-{a_{x'}\over \hbar \tilde{v}_0}\right]  |\bm E|^2 \Gamma_0\tilde{\tau}\\ && \hspace{7 mm}\times \left( {\tilde{\omega}_c\over\omega}\right)^2  \exp{\left[-\left({\beta \omega\over\tilde{\omega}_c }\right)^2\right]} F. \nonumber
\end{eqnarray}
Here the current is averaged over polarization assuming an unpolarized light,  and
\begin{eqnarray}
F&=&\sum_\pm \bigg[ f_0\left(\pm \tilde{E}_0+{\cal C} \text{sgn}(B_{x'}) a_{x'} k_\omega\right)\\
&&\mbox{}\hspace{3 mm} -f_0\left(\pm\tilde{E}_1+{\cal C} \text{sgn}(B_{x'}) a_{x'} k_\omega\right)\bigg],\nonumber
\end{eqnarray}
where
\begin{equation}
\tilde{E}_{0,1}={\hbar(\omega^2\mp\tilde{\omega}_c^2)\over2\omega},
\qquad
k_\omega={\tilde{E}_0\over \hbar\tilde{v}_0}.
\end{equation}

\subsubsection{Summation over monopoles in case of the $C_{2v}$ symmetry}

Equation~\eqref{j_tilt} demonstrates that, for $\bm B \parallel x'$, the MPGE current in each Weyl node  depends on  $a_{x'}$ and
\begin{equation}
a_{\perp}^2=a_{y'}^2+a_{z}^2.
\end{equation}
For the Weyl semimetals of $C_{2v}$ symmetry, the monopoles are located in four points of the Brillouine zone if they belong to the plane $z=0$ and in eight points if they are shifted from this plane. For simplicity we consider the former case. Let the first Weyl node be characterized by the chirality $\cal C$ and the tilt vector components be $a_{x'},a_{y'},a_{z}$.
The $C_2$ rotation  makes a transition to another monopole: 
\begin{equation}
	{\cal C}, a_{x'},a_{y'}^2+a_{z}^2 \to {\cal C}, -a_{x'}, a_{y'}^2+a_{z}^2.
\end{equation}
The mirror reflections $\sigma_v$  yield two other monopoles: 
\begin{equation}
	{\cal C}, a_{x'},a_{y'}^2+a_{z}^2 \to -{\cal C}, a_{y'},a_{x'}^2+a_{z}^2 \hspace{4 mm}
\end{equation}
and
\begin{equation}
{\cal C}, a_{x'},a_{y'}^2+a_{z'}^2 \to -{\cal C}, -a_{y'},a_{x'}^2+a_{z}^2.
\end{equation}

\begin{figure}[t]
\includegraphics[width=0.8\linewidth]{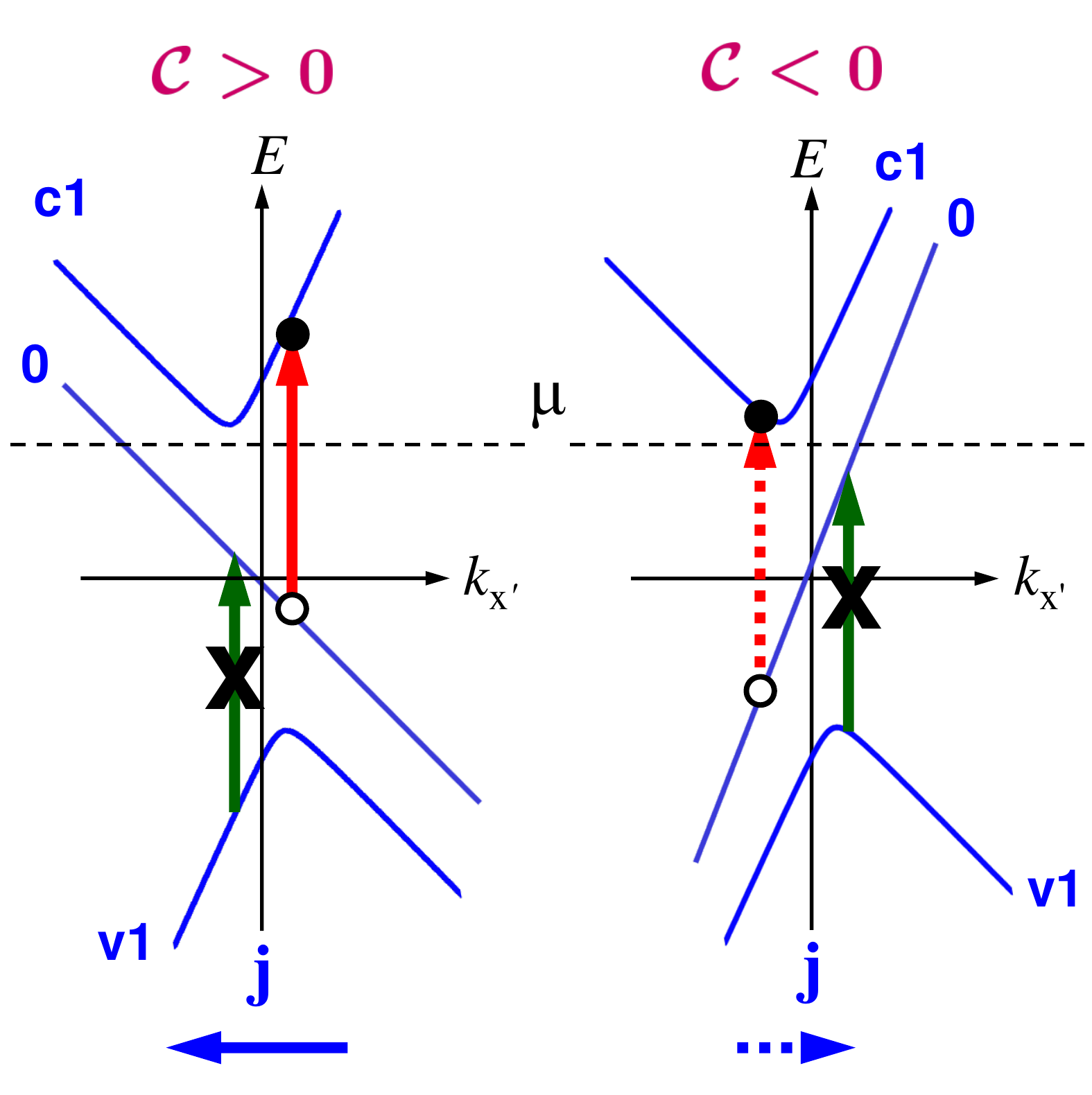} \qquad
\caption{Microscopic mechanism of the MPGE in a tilted Weyl semimetal. The net current is caused by a disbalance between contributions of the Weyl nodes of opposite chiralities: due to the tilt-dependent factor $\Phi$ defined by Eq.~(\ref{Phia}) the direct transition rate in the Weyl node with ${\cal C}>0$ becomes larger than that in the node with ${\cal C}<0$.}
\label{fig:j_10_tilt}
\end{figure}

Therefore, at zero temperature, there are two sources of the MPGE. The first mechanism is realized at the chemical potential lying below the bottom of the $c1$ subband, Fig.~\ref{fig:j_10_tilt}. In the first mechanism ${F=1}$, the photocurrent is dependent on $a^2_{\perp}$ via the parameters $\beta$ and $\gamma$~\cite{JETP_Lett_2017}, and the summation over four monopoles yields 
\begin{eqnarray}
j_{x'} &=& B_{x'}|\bm E|^2 {\cal C}  \Gamma_0\tilde{\tau} {2v_0^2e\over \hbar c \omega^2}\\ &&\times\left[\Phi(a_{y'}^2+a_{z}^2)-\Phi(a_{x'}^2+a_{z}^2)\right],\nonumber
\end{eqnarray}
where
\begin{equation} \label{Phia}
\Phi(a_{\perp}^2) = 2\gamma^3   \exp{\left[-{a_{\perp}^2\over \hbar^2 v_0^2 \gamma^3}\left({\omega\over\omega_c }\right)^2\right]}. 
\end{equation}
We remind that $\gamma=[1-(a_{\perp}/\hbar v_0)^2]^{-1/2}$. This equation demonstrates that the net MPGE current appears due to a difference in the direct optical transition rates in differently tilted Weyl nodes, Fig.~\ref{fig:j_10_tilt}.

The second mechanism is related with the interference of the $a_{x'}$-linear term in the velocity with the $a_{x'}$-dependent part of the distribution functions, Eq.~\eqref{j_tilt}. The photocurrent becomes nonzero due to a difference in the relaxation times $\tilde{\tau}$ and $\tau_p$. This mechanism working at quantized magnetic field is similar to that considered in Ref.~\cite{Kharzeev} for low magnetic fields.

\section{Discussion}
\label{disc}

\begin{figure}[t]
\includegraphics[width=0.9\linewidth]{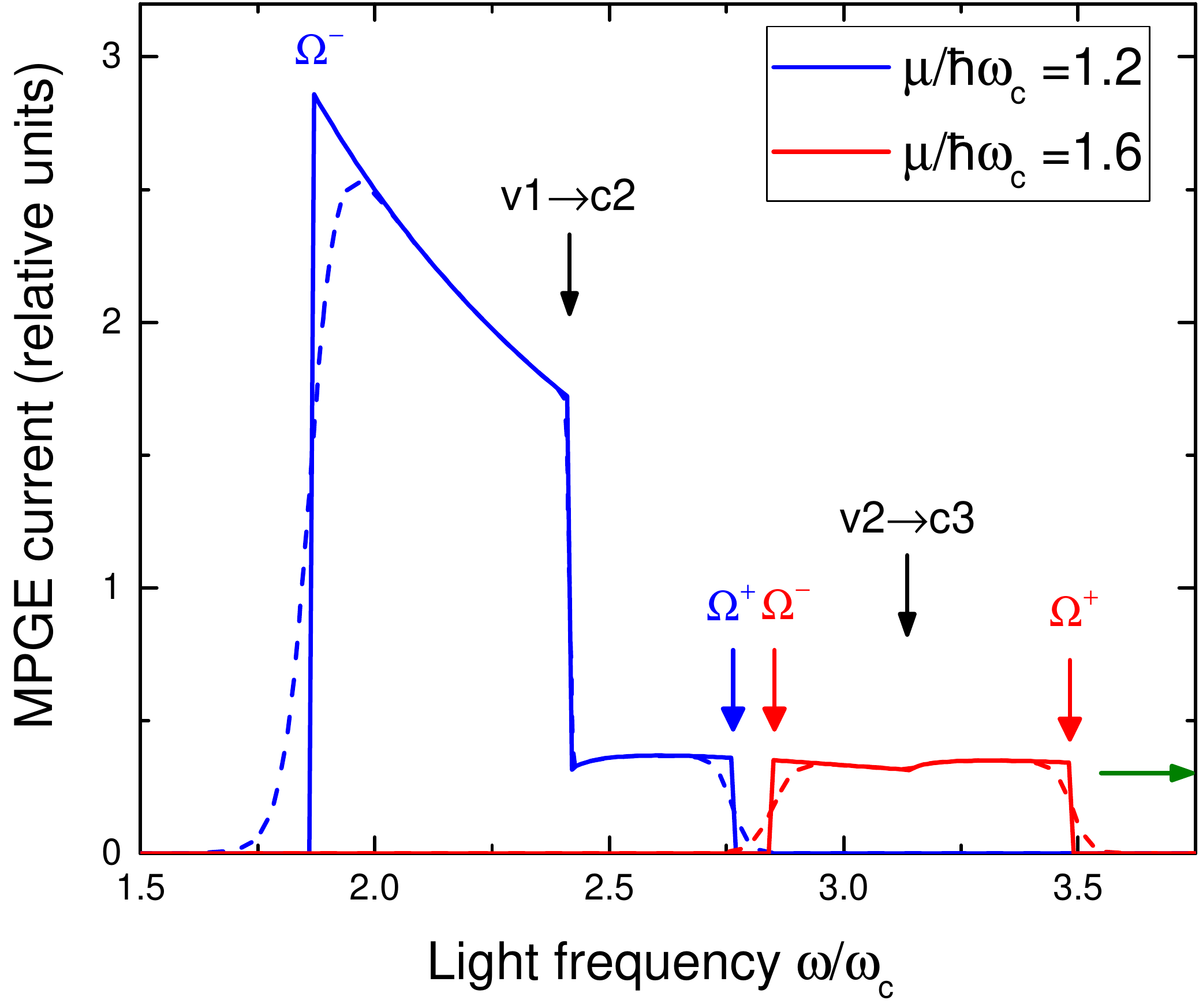}
\caption{ Frequency dependence of the MPGE current related to $|\bm E|^2 \Gamma_0\tau_p$ and calculated for two values of the chemical potential $\mu$. The relaxation times $\tilde{\tau}$ and $\tau_p$ are taken at a ratio of 10 to 1. The vertical arrows show the onset of the corresponding interband transition. The frequencies $\Omega^-(\mu)$ and $\Omega^+(\mu)$ defined by Eq.~(\ref{range_omega}) indicate the range where the photocurrent is induced at $T=0$. Solid and dashed curves are calculated for zero temperature and $T=0.01\hbar\omega_c$, respectively. The horizontal arrow shows the universal photocurrent value, Eq.~\eqref{MPGE_universal}.
At $\omega < \omega_c$ the photocurrent is also contributed by the intraband transitions  (not shown).}
\label{fig:j_MPGE}
\end{figure}

The calculated frequency dependence of the MPGE current is shown in Fig.~\ref{fig:j_MPGE}. In agreement with considerations of Sect.~\ref{MPGE_one_mode}, at zero temperature the photocurrent is induced in a limited interval of frequencies dependent on the level of chemical potential. For $\mu = 1.2 \hbar \omega_c < \sqrt{2} \hbar \omega_c$, the photocurrent  is controlled within the interval $\Omega_- < \omega < (\sqrt{2} + 1) \omega_c$ by the longer relaxation time $\tilde{\tau}$. On passing from Range~2 to Range~1 the photocurrent abruptly decreases because, although it is now contributed by the two transitions, $0 \to c1$ and $v1 \to c2$, the relaxation time shortens from  $\tilde{\tau}$ to $\tau_p$. For $\mu = 1.6 \hbar \omega_c > \sqrt{2} \hbar \omega_c$, the interband photocurrent  is generated under the transitions $0 \to c1$, $v1 \to c2$ and $v2 \to c3$ and controlled by the time $\tau_p$. 
In Fig.~\ref{fig:j_MPGE} dashed lines illustrate the effect of temperature on the photocurrent spectral dependence. At $\omega>\omega_c(\sqrt{2}+1)$, the photocurrent density is close to the universal value \eqref{MPGE_universal}.

It should be noted that, in addition to the circular photogalvanic photocurrent (\ref{general}), a transient electric current  can be generated under time-dependent optical excitation. Such a current appears due to a light-induced renormalization of the electron energy \cite{transient}. For the Weyl semimetal Hamiltonian (\ref{Hamilt3}) with a zero tilt term, the renormalized energy $\tilde{E}_{\pm, {\bm k}}$ differs from the unperturbed energy $E_{\pm, {\bm k}}$ by a correction $\delta E_{\pm, {\bm k}}$ which in the second-order approximation can be written as
\[
\delta E_{\pm, {\bm k}} = \pm \left\vert M_{+-}\right\vert^2 \frac{2 \hbar \omega}{(2 \hbar v_0 k)^2 - (\hbar \omega)^2}\:.
\]
The contribution (\ref{MOmega}) to the squared modulus of the matrix element $M_{+-}$ dependent on the circular polarization of the radiation is odd in the wavevector ${\bm k}$ and, therefore, the correction to the electron velocity ${\delta {\bm v}_{\pm, {\bm k}} = \hbar^{-1} \partial \delta E_{\pm, {\bm k}}/\partial {\bm k} }$ averaged over the direction of ${\bm k}$ does not vanish. As a result, at an abrupt switch-on of the light intensity at the moment $t=0$, a transient current 
$\delta {\bm j}_{\rm tr} (t) ={\rm e}^{- t/\tau_p} {\bm j}^0_{\rm tr}$ does appear, where
\[
 {\bm j}^0_{\rm tr} = \frac{e}{\hbar} \sum\limits_{\bm k} \frac{\partial \delta E_{+, {\bm k}} }{\partial {\bm k}}f_+(E_{+, {\bm k}})\:.
\]
The calculation performed at low temperature $T \approx 0$ for an $n$-doped sample with the Fermi wavevector satisfying the condition  $2 v_0k_\text{F} <  \omega$ results in 
\[
 {\bm j}^0_{\rm tr} =  \frac{w}{\pi} \frac{ {\bm j}_{\rm CPGE}}{\omega \tau_p}\:,
\]
where the circular current ${\bm j}_{\rm CPGE}$ is given by Eq.~(\ref{Gamma00}) and
\[
w = {\frac{2 \omega^2}{(2  v_0 k_\text{F})^2 -  \omega^2}}\:.
\]
It follows then that the non-stationary current $\delta {\bm j}_{\rm tr} (t)$, as compared with the photocurrent (\ref{Gamma00}), contains an additional small parameter $(\omega \tau_p)^{-1}$ and can be ignored, even in experiments with time-dependent light intensity.

Coulomb interaction effects on the direct optical absorption in systems with a linear dispersion has been investigated in graphene~\cite{Mishchenko,Schmalian_2009,JVH_2010,Teber_Kotikov_2014,Schmalian_2016}.
For three-dimensional WSMs, it has been shown that the electron-electron interaction yields a correction to the light absorption coefficient containing the factor $1/(N+1)$ which is small if $N \gg 1$, where $N$ is the number of Weyl points~\cite{WSM_opt_cond_2017,WSM_opt_cond_2018}. We remind that, for a Weyl semimetal of the $C_{4v}$ symmetry, $N=16$ for the nodes lying out of the plane $z=0$ and $N=8$ for the nodes on this plane.
The Coulomb correction to the CPGE current at interband transitions also contains the factor $1/(N+1)$. However, the analysis shows that the electron-electron effects are additionally suppressed in the CPGE current by a small factor $\ln^{-1}{(D/\omega)}$ where $D\gg\omega$ is the high-frequency cut-off.

\section{Conclusion}
\label{concl}
We have developed a theory of the circular and magneto-gyrotropic photogalvanic effects in Weyl semimetals with the point groups containing improper symmetry operations. In semimetals of the C$_{2v}$ symmetry with the linear energy dispersion, the net CPGE photocurrent becomes nonzero taking into account a spin-independent tilt term in the electron effective Hamiltonian. However, this is insufficient for the crystal class C$_{4v}$, like the TaAs Weyl semimetal. In this case one needs to add to the Hamiltonian not only the tilt but also spin-dependent terms of the second or third order in the electron quasi-momentum and take into account the nonlinear corrections, respectively, to the  velocity and Berry curvature in the equation for the current density. Additionally, the theory has been extended by consideration of the indirect intraband optical transitions and their contribution to the CPGE. Here an important point to bear in mind is that the probability rate is contributed by the composite (two-quantum) processes with virtual states both in the conduction and valence bands. 

Developing a theory of the polarization independent photocurrents in an external quantized magnetic field we have, in turn, analyzed three frequency ranges, namely, $\omega > \omega_c (\sqrt{2} + 1)$; $\omega_c (\sqrt{2} + 1) > \omega > \omega_c $ and
$\omega_c > \omega$ and found restrictions imposed by the level of chemical potential at zero temperature on the spectral intervals where the MPGE current is generated. The temperature smooths out the edges of these intervals. The calculation reveals that, for the C$_{2v}$ point-group symmetry, the net MPGE current is an even function of the tilt parameters $a_{\alpha}$.

\acknowledgements

We thank B.Z. Spivak for discussions. Financial support of the Russian Science Foundation (Project No. 17-12-01265) is acknowledged. Work of L.E.G. is supported by the Foundation for advancement of theoretical physics and mathematics ``BASIS''.

\end{document}